\documentclass{article}
\usepackage[top=2cm,bottom=3cm,left=2cm,right=2cm]{geometry}
\usepackage{amsmath}
\usepackage{amssymb} 
\usepackage{epsfig}
\usepackage{amsthm}
\usepackage{stmaryrd}
\usepackage{bm}
\usepackage{hyperref}
\usepackage{eufrak}
\usepackage{upgreek}
\usepackage{graphicx}

\newcommand{\inverttriangle}{%
               \mathrel{\raisebox{.1em}{%
               \reflectbox{\rotatebox[origin=c]{180}{$\triangle$}}}}}
               
\numberwithin{equation}{section}
\numberwithin{figure}{section}

\def\eq#1{(\ref{eq:#1})}
\def\lineup{\!\!\!\!\!\!\!\!\!\!&&}

\def\d{\partial}
\def\eps{\epsilon}

\def\fraction#1#2{ { \textstyle \frac{#1}{#2} }}

\def\deg{\mathrm{deg}}

\def\M{{\bf M}}
\def\m{{\bf m}}
\def\Q{{\bf Q}}
\def\n{{\bm \upeta}}

\def\b{{\bf b}}

\def\mmu{{\bm \upmu}}

\def\B{{\bf B}}
\def\bbeta{{\bm \upbeta}}
\def\C{{\bf C}}
\def\D{{\bf D}}

\def\H{{\bf \hat{H}}}
\def\G{{\bf \hat{G}}}
\def\F{{\bf \hat{F}}}
\def\U{{\bf \hat{U}}}
\def\V{{\bf \hat{V}}}
\def\E{{\bf \hat{E}}}

\def\PsiB{\Psi_B}
\def\PsiA{\Psi_{A}}
\def\Bnc{\tilde{B}^{\mathrm{noncyclic}}}

\newtheorem{Fact}{Property}

\begin{document}

\begin{titlepage}
\rightline{\tt LMU-ASC  32/15}
\rightline\today
\begin{center}
\vskip 2cm
\vskip 1.5cm {\large \bf{Relating Berkovits and $A_\infty$ Superstring Field Theories;\\ Small Hilbert Space Perspective}}
\vskip 1.0cm
{\large Theodore Erler}\footnote{email:tchovi@gmail.com}
\vskip 1.0cm
${}^1${\it {Arnold Sommerfeld Center, Ludwig-Maximilians University}}\\
{\it {Theresienstrasse 37, 80333 Munich, Germany}}\\

\vskip 2.0cm

{\bf Abstract}
\end{center}

In a previous paper it was shown that the recently constructed action for open superstring field theory based on $A_\infty$ algebras can be re-written in Wess-Zumino-Witten-like form, thus establishing its relation to Berkovits' open superstring field theory. In this paper we explain the relation between these two 
theories from a different perspective which emphasizes the small Hilbert space, and in particular the relation between the $A_\infty$ structures on both sides. 

\vskip 1.0cm
\noindent 

\noindent

\end{titlepage}

\tableofcontents

\section{Introduction}

Following the discovery of a new form of open superstring field theory based on $A_\infty$ algebras \cite{WittenSS}, a pressing question has been to understand its relation to the most established approach to open superstring field theory---namely, the Wess-Zumino-Witten-like theory proposed by Berkovits \cite{Berkovits1,Berkovits2}. In an accompanying paper \cite{OkWB} it was shown that the $A_\infty$ superstring field theory could be expressed in a Wess-Zumino-Witten-like form, thus providing a natural field redefinition relating it to Berkovits' superstring field theory. In this paper we explain the origin of this field redefinition from a different perspective which emphasizes the small Hilbert space.

The basic idea behind the construction of the $A_\infty$ superstring field theory is that string interactions can be generated by performing a kind of ``field redefinition" from a free theory. Usually field redefinitions do not effect $S$-matrix elements, but here the required ``field redefinition" is nontrivial since it is only defined in the large Hilbert space \cite{FMS}, and therefore is not consistent with the small Hilbert space constraint on the string field. One consequence of this construction is that the equations of motion of the $A_\infty$ theory can be written
\begin{equation}QG[\PsiA] = 0,\label{eq:introAinfEOM}\end{equation}
where $Q\equiv Q_B$ is the BRST operator, $\PsiA$ is the dynamical string field of the $A_\infty$ theory, and $G[\PsiA]$ is the ``field redefinition" which relates $\PsiA$ to a free field. Since $G[\Psi_A]$ is only defined in the large Hilbert space, we will call it an {\it improper} field redefinition.

Now consider Berkovits' superstring field theory. The equations of motion can be written in the form
\begin{equation}Q\Big((\eta e^{\Phi})e^{-\Phi}\Big)=0,\end{equation}
where $\eta\equiv\eta_0$ is the eta zero mode and $\Phi$ is the dynamical string field of the Berkovits theory. The field $\Phi$ is in the large Hilbert space. However, at the cost of fixing the part of the gauge invariance (associated with the $\eta$ zero mode), we can describe $\Phi$ in terms of a string field $\Psi_B$ in the small Hilbert space via \cite{INOT}
\begin{equation}\Phi = \xi\Psi_B,\end{equation}
where $\xi$ is an operator built from the $\xi$ ghost satisfying $[\eta,\xi]=1$. We refer to this partial gauge fixing as the {\it reduced} Berkovits theory. With this substitution, the equations of motion take the form
\begin{equation}Q\Big((\eta e^{\xi\Psi_B})e^{-\xi\Psi_B}\Big)=0,\end{equation}
Equivalently, we can write this as
\begin{equation}Q F[\PsiB] = 0\label{eq:introBerkEOM}\end{equation}
where 
\begin{equation}F[\PsiB ] = (\eta e^{\xi\PsiB})e^{-\xi\PsiB}\end{equation}
is an improper field redefinition which relates $\PsiB$ to a free field. Thus it is natural to suppose that the field redefinition between the $A_\infty$ and reduced Berkovits superstring field theories is given by equating the respective free fields:
\begin{equation}G[\PsiA] = F[\PsiB].\label{eq:GeqF}\end{equation}
As shown in \cite{OkWB} and as will be shown in the following, this turns out to be correct.

In contrast to \cite{OkWB}, however, this approach to relating the $A_\infty$ and Berkovits superstring field theories raises different kinds of questions. In particular, while the idea of generating interactions by ``field redefinition" is central to the $A_\infty$ superstring field theory, in Berkovits' superstring field theory this concept is novel and has not been articulated. Our approach to relating the Berkovits and $A_\infty$ theories aims to clarify this point. In particular, this informs our approach to the proof of equivalence of the actions, as well as our interest in understanding and comparing the $A_\infty$ structures of the two theories. 

This paper is organized as follows. Section \ref{sec:preliminaries} is devoted to mathematical background, in particular, characterizing the effect of field redefinitions on $A_\infty$ structures. This allows us, in section \ref{sec:G}, to describe in greater detail how the $A_\infty$ superstring field theory is defined by an improper field redefinition from a free theory. Then, in section \ref{sec:F}, we show that the reduced Berkovits theory can likewise be understood as an improper field redefinition of a free theory. The analysis, in particular, reveals that the reduced Berkovits theory can be characterized by a non-cyclic $A_\infty$ structure. However, non-cyclicity implies that the $A_\infty$ structure is not realized in the vertices of the reduced Berkovits action. In section \ref{sec:equivalence} we show that the $A_\infty$ and reduced Berkovits actions are equivalent by demonstrating that they can both be transformed to a free action when the coupling constant vanishes. We include a few appendices elucidating some technical aspects of the discussion. Appendix \ref{app:coproduct} introduces the coproduct for the purpose of clarifying the coalgebra representation of $A_\infty$ algebras.  Appendix \ref{app:nF} shows how the improper field redefinition of the reduced Berkovits theory transforms the $\eta$ zero mode. Appendix \ref{app:Bcyc} derives a formula relating the cyclic products in the reduced Berkovits action and the improper field redefinition of the reduced Berkovits theory. We end with some concluding remarks.

\section{Cohomomorphisms and Field Redefinitions}
\label{sec:preliminaries}

In this section we describe the coalgebra representation of $A_\infty$ algebras. Some of this is reviewed in recent papers \cite{WittenSS,ClosedSS,OkWB}, but we will need some further elaborations, especially about cohomomorphisms and their relation to field redefinitions. For a more mathematical discussion, see Kajiura \cite{Kajiura}. 

The central objects in our discussion are multi-string products
\begin{equation}c_n(A_1,...,A_n),\label{eq:cn}\end{equation}
where $A_1,...,A_n$ are string fields and the index on $c_n$ refers to the number of string fields which are being multiplied. It is useful to think of a multi-string product as a linear operator $c_n$ (denoted with the same symbol) which maps the tensor product of $n$ copies of the open string state space $\mathcal{H}$ into one copy:
\begin{equation}c_n: \mathcal{H}^{\otimes n}\to \mathcal{H}.\end{equation}
Acting on tensor products of states, the operator $c_n$ is defined
\begin{equation}c_n(A_1\otimes.... \otimes A_n) \equiv c_n(A_1,...,A_n), \label{eq:opprod}\end{equation}
where the right hand side is the multi-string product as denoted in \eq{cn}. Since tensor products of states can be used to form a basis, this defines $c_n$ on all of $\mathcal{H}^{\otimes n}$.

The formalism we need requires a different set of sign rules for (anti)commutation of operators and string fields than would be assigned by Grassmann parity. Therefore, we introduce a shifted grading on the open string state space called 
{\it degree}. The degree of a string field $A$, denoted $\deg(A)$, is defined to be Grassmann parity $\eps(A)$ plus one:\footnote{The change of grading from Grassmann parity to degree is often referred to as a {\it suspension} \cite{Tensor,MoellerSachs}.}
\begin{equation}\deg(A) = \eps(A)+1\ \ \ \ (\mathrm{mod}\ \mathbb{Z}_2).\label{eq:deg}\end{equation}
In particular, the field $\PsiB$ of the reduced Berkovits theory and the field $\PsiA$ of the $A_\infty$ theory are degree {\it even} (but Grassmann odd), and so should behave as {\it commuting} objects from the perspective of the new grading. Star multiplication, on the other hand, adds one unit of degree,
\begin{equation}\deg(A*B) = \deg(A)+\deg(B) + 1\ \ \ \ (\mathrm{mod}\ \mathbb{Z}_2),\end{equation}
and so should represent an {\it anticommuting} operation, like the BRST operator. However, the anticommuting nature of the star product is not immediately apparent. To see it, we must incorporate a sign factor into the definition of multiplication
\begin{equation}
m_2(A,B) \equiv (-1)^{\deg(A)} A*B.\label{eq:m2}
\end{equation}
With this extra sign, the derivation property of $Q$ and the associativity of the star product are reexpressed
\begin{eqnarray}
Qm_2(A,B) + m_2(QA,B) + (-1)^{\deg(A)}m_2(A,QB) \lineup = 0,\label{eq:Qder1}\\
m_2(m_2(A,B),C) + (-1)^{\deg(A)}m_2(A,m_2(B,C)) \lineup = 0.\label{eq:ass1}
\end{eqnarray}
Note that $Q$ acquires a minus sign when passing through $m_2$, as though $m_2$ were an anticommuting operation. Also, $Q$ acquires a sign when passing through the string field $A$ according to its {\it degree}, not its Grassmann parity. Ultimately, the different signs here boil down to a different choice of sign factors in the multi-string products which are ``natural" in the two grading schemes, as is the case for Witten's open string star product in \eq{m2}. We will mostly use degree as our grading, and multi-string products will be defined with the corresponding sign. See \cite{OkWB,Tensor,MoellerSachs} for more discussion. When appropriate, we will use the ordinary star product without the sign $A*B=AB$, especially in discussion of the Berkovits theory.

Often it is useful to express relations between multi-string products in the form of operator equations. Consider, in general, a pair of linear maps
\begin{equation} b_{k,m}: \mathcal{H}^{\otimes m}\to \mathcal{H}^{\otimes k},\ \ \ \ c_{\ell,n}: \mathcal{H}^{\otimes n}\to \mathcal{H}^{\otimes \ell}.
\end{equation}
We define a tensor product map
\begin{equation} 
b_{k,m}\otimes c_{\ell,n}: \mathcal{H}^{\otimes m+ n}\to \mathcal{H}^{\otimes k+\ell}
\end{equation}
as follows:
\begin{eqnarray}
\lineup (b_{k,m}\otimes c_{\ell,n})(A_1\otimes ... \otimes A_{m+n})\equiv\nonumber\\
\lineup \ \ \ \ \ \ \ \ \ \ \ \ \ \ 
 (-1)^{\deg(c_{\ell,n})(\deg(A_1)+....+\deg(A_m))} (b_{k,m}(A_1\otimes ...\otimes A_{m}))\otimes (c_{\ell,n}(A_{m+1}\otimes ... \otimes A_{m+n})).
 \ \ \ \ \ 
\end{eqnarray}
The degree of a linear map is defined as the degree of its output minus the sum of the degrees of its inputs, and the sign on the right hand side can be interpreted as arising from (anti)commuting $c_{n,\ell}$ past the first $m$ states.  With this we can rexpress the derivation property of $Q$ and associativity of the star product
\begin{eqnarray}
Qm_2 + m_2(Q\otimes \mathbb{I} + \mathbb{I}\otimes Q) \lineup = 0,\label{eq:Qder2}\\
m_2(m_2\otimes \mathbb{I} + \mathbb{I}\otimes m_2)\lineup = 0.\label{eq:ass2}
\end{eqnarray}
where $\mathbb{I}$ is the identity map on the open string state space. Acting the first equation on $A\otimes B$ and the second equation on $A\otimes B\otimes C$, and using \eq{opprod}, we recover the expected expressions \eq{Qder1} and \eq{ass1}.

\subsection{Coderivations and Cohomomorphisms}

When considering $A_\infty$ algebras and field redefinitions, we typically need to consider an infinite sequence of multi-string products acting on different numbers of open string states. To describe these objects in a unified setting, it is useful to consider the {\it tensor algebra} generated by taking tensor products of the open string state space:
\begin{equation}T\mathcal{H} = \mathcal{H}^{\otimes 0}\, \oplus\, \mathcal{H}\, \oplus\, \mathcal{H}^{\otimes 2}\, \oplus\, \mathcal{H}^{\otimes 3}\, \oplus\, ...\ .\end{equation}
Here $\mathcal{H}^{\otimes 0}$ is a 1-dimensional vector space given by multiplying a single basis vector $1_{T\mathcal{H}}$ by complex numbers. By definition, the basis vector $1_{T\mathcal{H}}$ is the identity of the tensor product, and so satisfies
\begin{equation}1_{T\mathcal{H}}\otimes V =V\otimes 1_{T\mathcal{H}} = V\end{equation}
for $V$ any element of the tensor algebra. The tensor algebra is a {\it coalgebra}, since there is a naturally defined operation called a {\it coproduct}, which produces an object in a pair of tensor algebras out of an object in one. This structure will not be essential for our considerations, but is sometimes useful, and provides some mathematical motivation for the definitions which follow. We give a more complete discussion in appendix \ref{app:coproduct}. 

We seek a way to promote an $n$-string product $D_n$ to a linear operator on the tensor algebra. This can be done by defining a so-called {\it coderivation} $\D_n$, corresponding to the product $D_n$. The operator $\D_n$ can be defined by its action on tensor products of states,
\begin{eqnarray}
\D_n A_1\otimes ... \otimes A_m \lineup = 0 
\ \ \ \ \ \ \ \ \ \ \ \ \ \ \ \ \ \ \ \ \ \ \ \ \ \ \ \ \ \ \ \ \ \ \ \ \ \ \ \ \ \ \ \ \ \ \ \ \ \ \ \ \ \ \ \ \ \ \mathrm{for}\ \ m<n,\\
\D_n A_1\otimes ... \otimes A_m \lineup = D_n (A_1, ... ,A_m)
\ \ \ \ \ \ \ \ \ \ \ \ \ \ \ \ \ \ \ \ \ \ \ \ \ \ \ \ \ \ \ \ \ \ \ \ \ \ \ \ \mathrm{for}\ \ m=n,\\
\D_n A_1\otimes ... \otimes A_m \lineup = \left(\sum_{\ell =0 }^{m-1} \mathbb{I}^{\otimes m-1-\ell}\otimes D_n\otimes\mathbb{I}^{\otimes \ell}\right)
 A_1\otimes ... \otimes A_m \ \ \ \ \mathrm{for}\ \ m\geq n,
\end{eqnarray}
where\footnote{$\mathbb{I}^{\otimes 0}$ is the identity operator on $\mathcal{H}^{\otimes 0}$. For any linear map $b_{k,m}$ from $\mathcal{H}^{\otimes m}$ to $\mathcal{H}^{\otimes k}$, it satisfies $\mathbb{I}^{\otimes 0}\otimes b_{k,m}=b_{k,m}\otimes \mathbb{I}^{\otimes 0} = b_{k,m}$.} 
\begin{equation}\mathbb{I}^{\otimes k}=\underbrace{\mathbb{I}\otimes ...\otimes\mathbb{I}}_{k\ \mathrm{times}}.\end{equation}
$\D_n$ acts as zero on $\mathcal{H}^{\otimes m}$ for $m<n$, but for $m\geq n$ it sums over all ways of multiplying $n$ sequential factors with $D_n$.
Let us introduce a projection operator $\pi_m$ from $T\mathcal{H}$ into $\mathcal{H}^{\otimes m}$. Then we can write this formula
 \begin{eqnarray}
\D_n\pi_m \lineup = 0,\ \ \ \ \mathrm{for}\ m<n,\\
\D_n\pi_m \lineup = \left(\sum_{\ell =0 }^{m-1} \mathbb{I}^{\otimes m-1-\ell}\otimes D_n\otimes\mathbb{I}^{\otimes \ell}\right)\pi_{m}\ \ \ \ \ \mathrm{for}\ m\geq n.\ \ \ \ \ \label{eq:Dpi}
\end{eqnarray}
If $\mathbb{I}_{T\mathcal{H}}$ is the identity operator on the tensor algebra, we have 
\begin{equation}\mathbb{I}_{T\mathcal{H}} = \sum_{m=0}^\infty \pi_m.\end{equation}
Acting $\D_n$ on both sides on this equation gives the formula:
\begin{equation}\D_n = \sum_{k=0}^{\infty}\left(\sum_{\ell =0 }^k \mathbb{I}^{\otimes k-\ell}\otimes D_n\otimes\mathbb{I}^{\otimes \ell}\right)\pi_{n+k}.\label{eq:Dn}\end{equation}
Another useful relation is
\begin{equation}\pi_{k+1} \D_n = \Big(D_n\otimes \underbrace{\mathbb{I}\otimes...\otimes\mathbb{I}}_{k\ \mathrm{t
imes}}\ +\ \mathbb{I}\otimes D_n\otimes\underbrace{\mathbb{I}\otimes...\otimes\mathbb{I}}_{k-1 \ \mathrm{times}}\ +\  ...\ +\ \underbrace{\mathbb{I}\otimes...\otimes\mathbb{I}}_{k\ \mathrm{times}}\otimes D_n\Big)\pi_{k+n}.\label{eq:piD}\end{equation}
In this paper we denote coderivations with a boldface. The degree of a coderivation is the same as the degree of its product. A general coderivation can be expressed as a sum of coderivations 
\begin{equation}\D = \D_0+\D_1+\D_2+\D_3+...\end{equation}
corresponding to a sequence of products\footnote{The zero string product can be seen as an operator which acts on $1_{T\mathcal{H}}$ and produces a string field. We write $D_0=D_0(1_{T\mathcal{H}})$.}
\begin{equation} D_0,\ \ D_1(A),\ \ D_2(A,B),\ \ D_3(A,B,C),\ \ ...\ .\end{equation}
Then in general we have
\begin{equation}\D = \sum_{\ell,m,n =0}^\infty(\mathbb{I}^{\otimes \ell}\otimes D_{m}\otimes \mathbb{I}^{\otimes n})\pi_{\ell+m+n}, \label{eq:D}\end{equation}
where the sum over $m$ runs over the sequence of products which defines $\D$. Note that the $n$th product in the coderivation $\D$ be can be recovered by computing $D_n = \pi_1\D\pi_n$.

A useful fact is that commutator of two coderivations (graded with respect to degree) is again a coderivation. Computing the commutator of $\C_m$ and $\D_n$, corresponding to the products $C_m$ and $D_n$, we find:\footnote{We always use $[\cdot,\cdot]$ to denote the commutator graded with respect to degree, except (mostly in the context of the Berkovits theory) when taking the commutator of string fields, which we define in the usual way with the star product $[A,B] \equiv A*B-(-1)^{\eps(A)}B*A$, and is graded with respect to Grassmann parity.}
\begin{eqnarray}
[\C_m,{\bf D}_n] \lineup = \C_m \D_n - (-1)^{\deg(C_m)\deg(D_n)}\D_n\C_m\nonumber\\
\lineup = \sum_{k=0}^{\infty}\left(\sum_{\ell =0 }^k \mathbb{I}^{\otimes k-\ell}\otimes [C_m,D_n]\otimes\mathbb{I}^{\otimes \ell}\right)\pi_{m+n-1+k},
\end{eqnarray}
where $[C_m,D_n]$ is an $m+n-1$ string product
\begin{eqnarray}[C_m,D_n] \lineup \equiv C_m \left(\sum_{\ell =0 }^{m-1} \mathbb{I}^{\otimes m-1-\ell}\otimes D_n\otimes\mathbb{I}^{\otimes \ell}\right)-D_n \left(\sum_{\ell =0 }^{n-1} \mathbb{I}^{\otimes n-1-\ell}\otimes C_m\otimes\mathbb{I}^{\otimes \ell}\right).
\label{eq:comprod}\end{eqnarray}
The commutator of coderivations $\C$ and $\D$, representing the sequences of products $C_0,C_1,C_2,...$ and $D_0,D_1,D_2,...$ respectively, is a coderivation $[\C,\D]$ representing the sequence of products
\begin{eqnarray}
\lineup [C_1,D_0]+[C_0,D_1],\nonumber\\
\lineup [C_2,D_0]+[C_1,D_1]+[C_0,D_2],\nonumber\\
\lineup [C_3,D_0]+[C_2,D_1]+[C_1,D_2]+[C_0,D_3],\nonumber\\
\lineup [C_4,D_0]+[C_3,D_1]+[C_2,D_2]+[C_1,D_3]+[C_0,D_4],\nonumber\\
\lineup\ \vdots \ .\label{eq:DEprod} \end{eqnarray}
Note that the expression $[C_0,D_0]$ would represent a ``$-1$ string product," and vanishes identically.

If we take the exponential of a coderivation of even degree, we obtain a linear operator on the tensor algebra called a {\it cohomomorphism}. A cohomomorphism is characterized by a sequence of degree even multi-string products 
\begin{equation}H_0,\ H_1(A),\ H_2(A,B),\ H_3(A,B,C),\ ...\end{equation}
packaged into an operator on the tensor algebra in the following way: 
\begin{eqnarray}
\H\lineup = \pi_0 + \sum_{\ell=1}^\infty\,\sum_{k_1,k_2,...,k_{\ell}=0}^\infty (H_{k_1}\otimes H_{k_2}\otimes ... \otimes H_{k_{\ell}})\pi_{k_1+k_2+...+k_{\ell}}.
\label{eq:H}
\end{eqnarray}
We always denote cohomomorphisms with a boldface and hat. In a moment we will see that coderivations of odd degree are closely related to $A_\infty$ algebras, while cohomomorphisms are related to field redefinitions. 

\subsection{$A_\infty$ Algebras}

An $A_\infty$ algebra is defined by a sequence of degree odd multi-string products
\begin{equation}D_0 = 0,\ \ D_1(A),\ \ D_2(A,B),\ \ D_3(A,B,C),\ ...\end{equation}
such that 
\begin{equation}\D = \D_1+\D_2+\D_3+...\end{equation}
is a nilpotent coderivation on the tensor algebra:
\begin{equation}[\D,\D]=0.\label{eq:D20}\end{equation}
It is conventional to assume that the zero string product $D_0$ vanishes. If it does not vanish, then the resulting algebraic structure is called a {\it weak} $A_\infty$ algebra, and appears when expanding string field theory around a state which does not satisfy the classical equations of motion \cite{Zwiebach}. From \eq{DEprod} the nilpotency of $\D$ is equivalent to the following identities, called $A_\infty$ relations:
\begin{equation}
0 =[D_1,D_1],\ \ \ 
0 = [D_1,D_2], \ \ \
0 = [D_1,D_3]+\frac{1}{2}[D_2,D_2],\ \ \ ... \ . \label{eq:Ainf}
\end{equation}
Acting on tensor products of states, the first three $A_\infty$ relations are
\begin{eqnarray}
0\lineup = D_1(D_1(A)),\\
0\lineup = D_1(D_2(A,B)) +D_2(D_1(A), B)+(-1)^{\deg(A)}D_2(A, D_1 (B)),\\
0\lineup = D_2(D_2(A,B),C) +(-1)^{\deg(A)} D_2(A,D_2(B,C))+ D_1(D_3(A,B,C))\nonumber \\
\lineup\ \ \ + D_3(D_1(A),B,C) + (-1)^{\deg(A)}D_3(A,D_1(B),C) + (-1)^{\deg(A)+\deg(B)}D_3(A,B,D_1(C)).
\end{eqnarray}
In string field theory, usually $D_1=Q$ is the BRST operator, and the first equation says that the BRST operator is nilpotent. The second equation says that $Q$ is a derivation of the 2-product $D_2$, and the third equation says that the associator of $D_2$ is the BRST variation of a 3-product $D_3$. 

The most basic example of an $A_\infty$ algebra is given by the BRST operator and the open string star product. Note that the nilpotency of $Q$, and equations \eq{Qder2} and \eq{ass2} can be written using \eq{comprod}
\begin{eqnarray}
0\lineup =[Q,Q],\\
0\lineup = [Q,m_2],\\
0\lineup = [m_2,m_2],
\end{eqnarray}
which implies that the coderivations
\begin{equation}\Q = \sum_{k=0}^{\infty}\left(\sum_{\ell =0 }^k \mathbb{I}^{\otimes k-\ell}\otimes Q\otimes\mathbb{I}^{\otimes \ell}\right)\pi_{k+1},
\ \ \ \ \ \m_2 = \sum_{k=0}^{\infty}\left(\sum_{\ell =0 }^k \mathbb{I}^{\otimes k-\ell}\otimes m_2\otimes\mathbb{I}^{\otimes \ell}\right)\pi_{k+2}
\end{equation}
satisfy
\begin{eqnarray}
0\lineup =[\Q,\Q],\\
0\lineup = [\Q,\m_2],\\
0\lineup = [\m_2,\m_2].
\end{eqnarray}
Note that the first and last equations mean that both $Q$ and $m_2$ separately define $A_\infty$ algebras, albeit ones with a single product. However, the middle equation implies that $\Q+\m_2$ defines an $A_\infty$ algebra: 
\begin{equation}[\Q+\m_2,\Q+\m_2] = 0.\end{equation}
This is the $A_\infty$ algebra relevant for open bosonic string field theory.

Incidentally, it is worth noting that this description of $A_\infty$ algebras is only possible if we use degree (rather than Grassmann parity) to measure the (anti)commutation of operators and string fields. To see why, note that if we had used the Grassmann grading, and defined multiplication $m_2(A,B) = A*B$ without the sign, associativity of the star product would be expressed
\begin{equation}m_2(m_2\otimes\mathbb{I}-\mathbb{I}\otimes m_2) = 0\ \ \ \ (\mathrm{Grassmann\ grading}).\end{equation}
This is not of the form $[m_2,m_2]$. In fact, $[m_2,m_2]$ would vanish identically because $m_2$ is Grassmann even.

\subsection{Field Redefinitions and $A_\infty$ Structures}
\label{subsec:FR}

Our goal is to relate two versions of open superstring field theory, and therefore a central question is how $A_\infty$ structures are changed by field redefinitions. Suppose we have an open string field theory described by an $A_\infty$ algebra $\D$. The equations of motion take the form
\begin{equation}0 = Q\Psi + D_2(\Psi,\Psi) + D_3(\Psi,\Psi,\Psi)+ ...\ .\label{eq:exEOM}\end{equation}
It is helpful to reexpress this in the tensor algebra. Given a string field $\Psi$, we can define a corresponding object in the tensor algebra called a {\it group-like element}:
\begin{equation}\frac{1}{1-\Psi}\equiv 1_{T\mathcal{H}}\,+\,\Psi\, + \,\Psi\otimes\Psi\, +\,\Psi\otimes\Psi\otimes\Psi\, +...=\sum_{n=0}^{\infty}\Psi^{\otimes n},\label{eq:gl}\end{equation}
where $\Psi^{\otimes 0}=1_{T\mathcal{H}}$. Note that
\begin{eqnarray}
\pi_1 \D\frac{1}{1-\Psi} \lineup = \pi_1\Big(\Q+\D_2+\D_3+...\Big)\Big(1_{T\mathcal{H}}\,+\,\Psi\, + \,\Psi\otimes\Psi\, +\,\Psi\otimes\Psi\otimes\Psi\, +...\Big),\nonumber\\
\lineup = \Big(Q\pi_1+D_2\pi_2+D_3\pi_3+...\Big)\Big(1_{T\mathcal{H}}\,+\,\Psi\, + \,\Psi\otimes\Psi\, +\,\Psi\otimes\Psi\otimes\Psi\, +...\Big),
\nonumber\\
\lineup =  Q\Psi + D_2(\Psi,\Psi) + D_3(\Psi,\Psi,\Psi)+ ...\ , \label{eq:piEOM1}
\end{eqnarray}
where we expanded $\D$ out into coderivations representing its component products and used \eq{piD}. Therefore the equations of motion can be written
\begin{equation}\pi_1 \D\frac{1}{1-\Psi}=0.\end{equation}
We can simplify further. Note that the action of $\D$ on a group-like element takes the form
\begin{eqnarray}
\D\frac{1}{1-\Psi}\lineup = \sum_{\ell,m,n =0}^\infty\Psi^{\otimes \ell}\otimes D_{m}(\underbrace{\Psi,...,\Psi}_{m\ \mathrm{times}})
\otimes \Psi^{\otimes n},\\
\lineup = \frac{1}{1-\Psi}\otimes \left(\pi_1 \D\frac{1}{1-\Psi}\right)\otimes\frac{1}{1-\Psi}.
\end{eqnarray}
Therefore the equations of motion are equivalent to
\begin{equation}\D\frac{1}{1-\Psi}=0.\label{eq:EOM0}\end{equation}
Now we want to see what happens when we make a field redefinition
\begin{equation}\Psi = H[\Psi'], \end{equation}
where $H[\Psi']$ is a functional of a new degree even string field $\Psi'$. Making a power series expansion in $\Psi'$, we can read off a sequence 
of degree even multi-string products:
\begin{equation}H[\Psi']= H_0+H_1(\Psi') + H_2(\Psi',\Psi') + H_3(\Psi',\Psi',\Psi')+...\ .\end{equation}
The zero-string product $H_0$ represents a vacuum shift, while the $1$-string product represents a linear transformation of the string field. The the higher products are only defined up to terms which vanish when the states being multiplied are equal, but for our purposes it will not matter how this ambiguity is fixed. By a similar argument as in \eq{piEOM1}, we can write the field redefinition
\begin{equation}H[\Psi'] = \pi_1\H\frac{1}{1-\Psi'},\end{equation}
where $\H$ is the cohomomorphism corresponding to the products $H_n$. Let's see what happens to the group-like element under field redefinition: 
\begin{eqnarray}
\frac{1}{1-\Psi}\lineup = \frac{1}{1-H[\Psi']},\\
\lineup = \sum_{\ell = 0}^\infty H[\Psi']^{\otimes \ell},\\
\lineup = 1_{T\mathcal{H}}+\sum_{\ell = 1}^\infty \, \sum_{k_1,..., k_{\ell}=0}^{\infty}H_{k_1}(\underbrace{\Psi',...,\Psi'}_{k_1\ \mathrm{times}})\otimes ... \otimes H_{k_{\ell+1}}( \underbrace{\Psi',...,\Psi'}_{k_{\ell}\ \mathrm{times}}),\\
\lineup = \H \frac{1}{1-\Psi'}.
\end{eqnarray} 
The equations of motion for $\Psi'$ can therefore be expressed
\begin{equation}0=\D\H\frac{1}{1-\Psi'}.\label{eq:pEOM1}\end{equation}
Since the field redefinition $H[\Psi]$ is invertible, there is an inverse cohomomorphism $\H^{-1}$ satisfying 
\begin{equation}\H^{-1}\H = \H\H^{-1} = \mathbb{I}_{T\mathcal{H}}.\end{equation}
The inverse field redefinition is simply 
\begin{equation}\Psi'= H^{-1}[\Psi] = \pi_1 \H^{-1}\frac{1}{1-\Psi}, \end{equation}
as can be checked:
\begin{equation}
H^{-1}[H[\Psi']] = \pi_1 \H^{-1}\frac{1}{1-H[\Psi']} = \pi_1 \H^{-1}\H\frac{1}{1-\Psi'} = \pi_1\frac{1}{1-\Psi'} = \Psi'.
\end{equation}
Now if we multiply the equations of motion in \eq{pEOM1} by $\H^{-1}$ we find the operator
\begin{equation}\D' = \H^{-1}\D\H,\end{equation}
and
\begin{equation}\D' \frac{1}{1-\Psi'}=0.\label{eq:pEOM2}\end{equation}
The operator $\D'$ is a coderivation (see appendix A). Moreover, it is nilpotent
\begin{equation}[\D',\D']=0,\end{equation}
as follows from the nilpotency of $\D$. The zero-string product of $\D'$ can be derived by acting on the zero-string component of the tensor algebra:
\begin{eqnarray}D_0' \lineup = \D' 1_{T\mathcal{H}}, \\
\lineup = \H^{-1} \D\H \, 1_{T\mathcal{H}},\\
\lineup = \H^{-1}\D\frac{1}{1-H_0},\end{eqnarray}
where we use the fact that $\H\, 1_{T\mathcal{H}}$ is a group-like element corresponding to the zero string product $H_0$. Provided $H_0$ shifts to a vacuum which satisfies the equations of motion, $D_0'$ vanishes. Therefore $\D'$ describes an $A_\infty$ algebra, and in particular the equations of motion of 
$\Psi'$ are 
\begin{equation}0= D_1'(\Psi') + D_2'(\Psi',\Psi') + D_3'(\Psi',\Psi',\Psi')+ ...\ .\end{equation}
where the new products $D_n'$ satisfy $A_\infty$ relations. In summary, field redefinitions map between $A_\infty$ algebras provided they map solutions into solutions. 

\subsection{Symplectic Form and Cyclicity}

To find an action principle we need a nondegenerate inner product between open string states. For NS open superstring field theory, this is provided by the BPZ inner product.  We will need both large and small Hilbert space BPZ inner products, which we denote respectively with a subscript $L$ and a subscript $S$:
\begin{eqnarray}
\langle A,B\rangle_L \lineup = \langle I\circ A(0) B(0)\rangle^{\mathrm{UHP}}_L,\label{eq:largeBPZ}\\
\langle A,B\rangle_S \lineup= \langle I\circ A(0) B(0)\rangle^{\mathrm{UHP}}_S,\label{eq:smallBPZ}
\end{eqnarray}
where, in the second equation $A$ and $B$ are assumed to be in the small Hilbert space, i.e. $\eta A = \eta B=0$. On the right hand side of these equations, 
$A(0)$ and $B(0)$ are the vertex operators corresponding to the states $A$ and $B$, and $I(z) = -1/z$ is the BPZ conformal map, and the correlators are computed on the upper half plane and in the large/small Hilbert spaces, respectively, with normalization
\begin{eqnarray}
\langle \xi c\d c\d^2 c e^{-2\phi}(0)\rangle_L^{\mathrm{UHP}} \lineup = 2,\\
\langle c\d c\d^2 c e^{-2\phi}(0)\rangle_S^{\mathrm{UHP}} \lineup = -2.
\end{eqnarray}
In the large Hilbert space, the BPZ inner product is nonvanishing on states whose ghost number adds to 2 and whose picture number adds to $-1$. In the small Hilbert space, the BPZ inner product is nonvanishing on states whose ghost number adds to 3 and whose picture number adds to $-2$. Since we are grading the open string state space with respect to degree, it is convenient to include an extra sign factor into the inner product, defining
\begin{eqnarray}\omega_L(A,B) \lineup = (-1)^{\deg(A)}\langle A,B\rangle_L,\\
\omega_S(A,B) \lineup = (-1)^{\deg(A)}\langle A,B\rangle_S.
\end{eqnarray}
The following properties of $\omega(A,B)$ hold in both the large and small Hilbert space, so we will temporarily omit the subscript. The usual symmetry of the BPZ inner product implies that $\omega$ is graded antisymmetric with respect to degree
\begin{equation}\omega(A,B) = -(-1)^{\deg(A)\deg(B)} \omega(B,A),\end{equation}
and therefore is a {\it symplectic form}. By definition, an $n$-string product $c_n$ is {\it cyclic} if it satisfies  
\begin{equation}\omega(\Psi_1,c_n(\Psi_2,...,\Psi_{n+1})) = -(-1)^{\deg(c_n)\deg(\Psi_1)}\omega(c_n(\Psi_1,...,\Psi_n),\Psi_{n+1}).\end{equation}
When this property is satisfied, antisymmetry of the symplectic form implies that the vertex is invariant under cyclic permutations of the inputs (up to the appropriate signs from anticommutation). In the spirit of \cite{GJ}, we can describe the symplectic form as a 2-string vertex which maps two copies of the state space into a complex number:
\begin{equation}\langle \omega|: \mathcal{H}\otimes\mathcal{H}\to\mathbb{C},\end{equation}
so that 
\begin{equation}\langle \omega |A\otimes B = \omega(A,B).\end{equation}
A cyclic product $c_n$ then satisfies
\begin{equation}\langle \omega| \mathbb{I}\otimes c_n = -\langle\omega | c_n\otimes\mathbb{I}.\end{equation}
In general, multi-string products are not cyclic. We can define a cyclically permuted product $\omega\circ c_n$ through the formula,
\begin{equation}\langle\omega | \mathbb{I}\otimes c_n \equiv -\langle \omega| (\omega\circ c_n) \otimes \mathbb{I},\end{equation}
so that $\omega\circ c_n = c_n$ if the product is cyclic. In general we can consider a $k$-fold cyclic permutation of a product
\begin{equation}\omega^{k}\circ c_n = \underbrace{\omega\circ (... \circ (\omega\circ}_{k\ \mathrm{times}} c_n)...).\end{equation}
Since an $(n+1)$-string vertex can only be cyclically permuted $n$ times before coming back to itself, we have $\omega^{n+1}\circ c_n = c_n$.

An $A_\infty$ algebra $\D$ whose multi-string products are cyclic $\omega\circ D_n = D_n$ is called a {\it cyclic} $A_\infty$ algebra. Most 
$A_\infty$ algebras considered in string field theory are cyclic, but this turns out not to be the case in the reduced Berkovits theory.

\section{Generating the $A_\infty$ Theory from a Free Theory}
\label{sec:G}

In this section we review how the $A_\infty$ superstring field theory can be generated by an improper field redefinition of a free theory. The improper field redefinition appears on the left-hand side of the equation
\begin{equation}G[\PsiA]=F[\PsiB],\end{equation}
which relates the $A_\infty$ superstring field theory and the reduced Berkovits superstring field theory. 

The $A_\infty$ superstring field theory governs the dynamics of a degree even NS string field $\PsiA$ which carries ghost number $1$ and picture number $-1$, and lives in the small Hilbert space:
\begin{equation}\eta\PsiA = 0.\end{equation}
The action takes the form
\begin{equation}S_{A} = \frac{1}{2}\omega_S(\PsiA,Q\PsiA) + \frac{1}{3}\omega_S(\PsiA,M_2(\PsiA,\PsiA))+
\frac{1}{4}\omega_S(\PsiA,M_3(\PsiA,\PsiA,\PsiA))+ ...\ ,\label{eq:SA}\end{equation}
where $Q,M_2,M_3,...$ are a sequence multi-string products of odd degree which satisfy the relations of a cyclic $A_\infty$ algebra. The products are in the small Hilbert space
\begin{equation}[\eta,M_n] = 0,\end{equation}
and the product $M_{n+1}$ carries picture number $n$ and ghost number $1-n$ (otherwise the corresponding vertex would vanish).  For example, the 2-string product is defined
\begin{equation}M_2(A,B) = \frac{1}{3}\Big[X m_2(A,B) + m_2(XA,B)+m_2(A,XB) \Big],\label{eq:M2}\end{equation}
or, factoring out $A\otimes B$
\begin{equation}M_2= \frac{1}{3}\Big[X m_2+ m_2(X\otimes \mathbb{I}+ \mathbb{I}\otimes X)\Big].\end{equation}
Here, $m_2$ is the open string star product (including the sign of \eq{m2}) and the picture-changing operator $X$ is defined 
\begin{equation}X= [Q, \xi],\end{equation}
where $\xi$ is an operator built from the $\xi$ ghost:
\begin{equation}\xi = \oint_{|z|=1}\frac{dz}{2\pi i}f(z)\xi(z).\label{eq:xi}\end{equation}
We assume that the function $f(z)$ is a analytic in the neighborhood of the unit circle and satisfies two additional properties:
\begin{eqnarray}
\lineup \mathrm{i)}\ \ \ \ \ f(z) = -\frac{1}{z^2}f\left(-\frac{1}{z}\right),\\
\lineup \mathrm{ii)}\ \ \ \ \oint_{|z|=1}\frac{dz}{2\pi i} f(z) = g_o.
\end{eqnarray}
The first property implies that $\xi$ is BPZ even:
\begin{equation} \langle \omega_L| \mathbb{I}\otimes \xi = \langle \omega_L|\xi\otimes\mathbb{I},\end{equation}
and the second implies that $\xi$ anticommutes with $\eta$ to give the open string coupling constant:
\begin{equation}\ [\eta,\xi] = g_o.\end{equation}
The fact that the commutator $[\eta,\xi]$ can be identified with the coupling constant is not immediately obvious, but it can be explained as follows. From ghost and picture number counting, it is clear that the $(n+2)$-string product $M_{n+2}$ must contain one insertion of $X$ and $n$ insertions of $\xi$. Now assuming $g_o\neq 0$ we can write
\begin{equation}\xi = g_o\xi',\end{equation}
where $[\eta,\xi']=1$. Since the total number of $\xi$s plus $X$s is equal $n+1$, we can pull $n+1$ powers of $g_o$ out of $M_{n+2}$ while replacing $\xi$ with $\xi'$: 
\begin{equation}M_{n+2} = g_o^{n+1}(M_{n+2})|_{\xi'}.\end{equation}
The action becomes
\begin{equation}S_{A} = \frac{1}{2}\omega_S(\PsiA,Q\PsiA) + \frac{g_o}{3}\omega_S(\PsiA,M_2(\PsiA,\PsiA)|_{\xi'})+
\frac{g_o^2}{4}\omega_S(\PsiA,M_3(\PsiA,\PsiA,\PsiA)|_{\xi'})+ ...\ .\end{equation}
where it is clear that we can identify $g_o$ with the open string coupling constant. 

Let us demonstrate---up to cubic order and when $g_o=0$---that the $A_\infty$ superstring field theory can be obtained by field redefinition from a free theory. Take the free action
\begin{equation}S_\mathrm{free} = \frac{1}{2}\omega_S(\Psi_0,Q\Psi_0),\end{equation}
and plug in the field redefinition 
\begin{equation}\Psi_0 = \PsiA + \mu_2(\PsiA,\PsiA) + \mathrm{higher\ orders},\label{eq:FR2}\end{equation}
where $\mu_2$ is a degree even 2-product of the form
\begin{equation}\mu_2= \frac{1}{3}\Big[\xi m_2- m_2(\xi\otimes \mathbb{I}+ \mathbb{I}\otimes \xi)\Big].\label{eq:mu2}\end{equation}
Here we must take $g_o=0$ otherwise the field redefinition cannot be defined in the small Hilbert space. Note that $\mu_2$ is cyclic:
\begin{equation}\langle \omega_L|\mathbb{I}\otimes\mu_2 = -\langle \omega_L|\mu_2\otimes\mathbb{I} .\end{equation}
The action becomes
\begin{eqnarray}
S_\mathrm{free}\lineup = \frac{1}{2}\omega_S(\PsiA,Q\PsiA)+\frac{1}{2}\omega_S(\PsiA,Q\mu_2(\PsiA,\PsiA)) + \frac{1}{2}\omega_S(\mu_2(\PsiA,\PsiA),Q\PsiA)+\mathrm{higher\ orders},\nonumber\\
\lineup = \frac{1}{2}\omega_S(\PsiA,Q\PsiA)+\omega_S(\PsiA,Q\mu_2(\PsiA,\PsiA)) +\mathrm{higher\ orders},\nonumber\\
\lineup = \frac{1}{2}\omega_S(\PsiA,Q\PsiA)+\frac{1}{3}\omega_S\Big(\PsiA,Q\mu_2(\PsiA,\PsiA) - \mu_2(Q\PsiA,\PsiA) - \mu_2(\PsiA,Q\PsiA)\Big) 
+\mathrm{higher\ orders},\nonumber\\
\label{eq:SfSA}
\end{eqnarray}
where in the third step we used cyclicity of $\mu_2$ to distribute $Q$ symmetrically on each entry of the vertex. The product in the cubic term is BRST exact: 
\begin{equation}Q\mu_2 -\mu_2(Q\otimes\mathbb{I} + \mathbb{I}\otimes Q) =[Q,\mu_2].\label{eq:M2mu2}\end{equation}
Computing the BRST commutator replaces the $\xi$ insertion in $\mu_2$ with $X$, and we find
\begin{equation}M_2=[Q,\mu_2].\end{equation} 
The action can be written
\begin{equation} 
S_\mathrm{free} = \frac{1}{2}\omega_S(\PsiA,Q\PsiA) + \frac{1}{3}\omega_S(\PsiA,M_2(\PsiA,\PsiA))+\mathrm{higher\ orders}.
\end{equation}
Therefore---at least to this order and when $g_o=0$---the action for the $A_\infty$ superstring field theory is generated from a free action by field redefinition. 

When $g_o\neq 0$ the $A_\infty$ theory is not a free theory. Still, the 2-product $M_2 = [Q,\mu_2]$ appears as though it is generated from a free theory by field redefinition. This generalizes to the higher products as follows. The products $Q,M_2,M_3,M_4...$ can be packaged into a coderivation $\M$
\begin{equation}\M = \Q +\M_2+\M_3+\M_4+...\ \end{equation}
which takes the form
\begin{equation}\M = \G^{-1}\Q \G, \label{eq:MG1}\end{equation}
where $\G$ is a cohomomorphism whose form we will review shortly. Recalling section \ref{subsec:FR}, it is clear that $\M$ can be derived from free equations of motion
\begin{equation}Q\Psi_0 = 0,\end{equation}
upon substituting
\begin{eqnarray}\Psi_0 \lineup  = \pi_1 \G\frac{1}{1-\PsiA}=G[\PsiA].\end{eqnarray}
Since $\G$ is in the large Hilbert space when $g_o\neq 0$, this is an improper field redefinition. This is why the $A_\infty$ superstring field theory is a nontrivial interacting theory.

\subsection{Construction of Multi-string Products}

To define $\G$ we must review the construction of the multi-string products given in \cite{WittenSS}. The products $M_{n+1}$ are defined by a set of recursive equations involving three sequences of multi-string products, which we refer to as follows:\footnote{In \cite{WittenSS}, the $\mu_n$s are referred to as "dressed" products and denoted $\overline{M}_{n+2}$. The current terminology follows \cite{ClosedSS}.}
\begin{eqnarray}
M_{n+1} \lineup = \mathrm{products},\ \ \ \ \ \ \ \ \ \ \ (\mathrm{degree\ odd,\ picture}\ n),\\
m_{n+2}\lineup = \mathrm{bare\ products} \ \ \ \ \ \ (\mathrm{degree\ odd,\ picture}\ n),\\
\mu_{n+2} \lineup = \mathrm{gauge\ products}\ \ \ \ (\mathrm{degree\ even,\ picture}\ n+1).
\end{eqnarray}
The products $M_{n+1}$ start with $M_1=Q$, while the bare products and gauge products start with 2-string multiplication $m_2$ and $\mu_2$. The recursive equations for these products are described by promoting $M_{n+1},\mu_{n+2}$ and $m_{n+2}$ to coderivations on the tensor algebra and defining generating functions
\begin{eqnarray}
\M(t)\lineup = \sum_{n=0}^\infty t^n \M_{n+1},\\
\m(t)\lineup = \sum_{n=0}^\infty t^n \m_{n+2},\\
\mmu(t) \lineup = \sum_{n=0}^\infty t^n \mmu_{n+2}.
\end{eqnarray}
Note that at $t=1$ we have $\M(1) = \M$. The generating functions are postulated to satisfy the differential equations
\begin{eqnarray}
\frac{d}{dt}\M(t) \lineup = [\M(t),\mmu(t)],\label{eq:rec1}\\
\frac{d}{dt}\m(t) \lineup = [\m(t),\mmu(t)],\label{eq:rec3}\\
\ [\n,\mmu(t)] \lineup = g_o \,\m(t).\label{eq:rec2}
\end{eqnarray}
Expanding in powers of $t$, we find a set of recursive equations which define the higher products in terms of commutators of lower ones. See \cite{WittenSS} for a more detailed description of the recursion. Note that the solution to the last equation \eq{rec2} is not unique, since we can always add an $\eta$-exact term to $\mu_{n+2}$.  As we will see, this choice is ultimately what distinguishes the $A_\infty$ superstring field theory from the reduced Berkovits superstring field theory. The $A_\infty$ theory makes a specific choice:
\begin{equation}\mu_{n+2} = \frac{1}{n+3} \Big(\xi m_{n+2} - m_{n+2}(\xi\otimes\underbrace{\mathbb{I}\otimes...\otimes\mathbb{I}}_{n+1\ \mathrm{times}} + \mathbb{I}\otimes\xi\otimes\underbrace{\mathbb{I}\otimes...\otimes\mathbb{I}}_{n\ \mathrm{times}}+ ... + 
\underbrace{\mathbb{I}\otimes...\otimes\mathbb{I}}_{n+1\ \mathrm{times}}\otimes \xi)\Big).\label{eq:rec2inv}\end{equation}
This generalizes \eq{mu2}, and guarantees that the products generated by the recursion are cyclic \cite{WittenSS}.

\subsection{Construction of Improper Field Redefinition}

Consider the cohomomorphism\footnote{The fact that this is a cohomomorphism is explained in appendix \ref{app:coproduct}}
\begin{equation}\G(t_1,t_2) = \mathcal{P}\left[\exp\left(\int_{t_1}^{t_2} dt\, \mmu(t)\right)\right]\label{eq:Gt1t2}\end{equation}
where the path ordered exponential is defined in sequence of increasing $t$. We can express the generating functions in the form
\begin{eqnarray}
\mmu(t) \lineup = \G(0,t)^{-1}\frac{d}{dt} \G(0,t),\label{eq:mmuG}\\
\M(t) \lineup = \G(0,t)^{-1}\Q\G(0,t),\label{eq:MG}\\
\m(t) \lineup = \G(0,t)^{-1}\m_2 \G(0,t).\label{eq:mG}
\end{eqnarray}
The first equation follows from the definition of the path ordered exponential. The second two equations follow because they provide a solution to the differential equations \eq{rec1} and \eq{rec3} with the correct initial conditions, namely $\M(0) = \Q$ and $\m(0) = \m_2$. In particular if 
\begin{equation}
\G \equiv \G(0,1),\label{eq:G}
\end{equation}
we obtain 
\begin{equation}\M = \G^{-1}\Q\G.\end{equation}
as anticipated in \eq{MG1}. Now we can compute the improper field redefinition which produces the $A_\infty$ theory from a free theory. Keeping track of terms out to third order in the field, we have explicitly
\begin{eqnarray}
\lineup\!\!\!\!\!\!\!\!\!\!\! G[\PsiA ] = \pi_1\G\frac{1}{1-\PsiA}\nonumber\\
\lineup \!\!\!\!\!\!\!\!\!\!\!=\pi_1\left(\mathbb{I}_{T\mathcal{H}}+ \int_0^1\!\! dt\, \mmu(t) + \int_0^1\!\! dt_1\int_{t_1}^1\!\! dt_2 \,\mmu(t_1)\mmu(t_2) + ...\right)\Big(1_{T\mathcal{H}}+\PsiA 
+ \PsiA\!\otimes\!\PsiA + \PsiA\!\otimes\!\PsiA\!\otimes\!\PsiA + ...\Big),\nonumber\\
\lineup\!\!\!\!\!\!\!\!\!\!\! = \pi_1\left(\mathbb{I}_{T\mathcal{H}}+\mmu_2+\frac{1}{2}\mmu_3 +\frac{1}{2}\mmu_2^2 + ... \right)\Big(1_{T\mathcal{H}}+\PsiA 
+ \PsiA\otimes\PsiA + \PsiA\otimes\PsiA\otimes\PsiA + ...\Big)\nonumber\\
\lineup\!\!\!\!\!\!\!\!\!\!\! = \left(\pi_1+\mu_2\pi_2 +\frac{1}{2}\mu_3\pi_3 +\frac{1}{2}\mu_2\pi_2\mmu_2 + ... \right)\Big(1_{T\mathcal{H}}+\PsiA 
+ \PsiA\otimes\PsiA + \PsiA\otimes\PsiA\otimes\PsiA + ...\Big),\nonumber\\
\lineup \!\!\!\!\!\!\!\!\!\!\!
= \left(\pi_1+\mu_2\pi_2+\frac{1}{2}\mu_3\pi_3 +\frac{1}{2}\mu_2(\mu_2\otimes\mathbb{I}+\mathbb{I}\otimes\mu_2)\pi_3 + ... \right)
\Big(1_{T\mathcal{H}}+\PsiA 
+ \PsiA\!\otimes\!\PsiA + \PsiA\!\otimes\!\PsiA\!\otimes\!\PsiA + ...\Big),\nonumber\\
\lineup\!\!\!\!\!\!\!\!\!\!\! = \PsiA + \mu_2(\PsiA,\PsiA) + \frac{1}{2}\Big(\mu_3(\PsiA,\PsiA,\PsiA) +\mu_2(\PsiA,\mu_2(\PsiA,\PsiA)) + \mu_2(\mu_2(\PsiA,\PsiA),\PsiA)\Big)+...\ . \label{eq:Gexp}
\end{eqnarray}
Here we expanded $\G$ in terms of $\mmu(t)$, substituted the expression for $\mmu(t)$ as a generating function for $\mmu_{n+2}$, performed the integration over $t$, and used \eq{piD}. For the sake of being explicit, the gauge 3-product appearing above is
\begin{equation}
\mu_3 = \frac{1}{4}\Big(\xi m_3 + m_3(\xi\otimes\mathbb{I}\otimes\mathbb{I} + \mathbb{I}\otimes\xi\otimes\mathbb{I} + \mathbb{I}\otimes\mathbb{I}\otimes\xi)\Big),\label{eq:mu3}
\end{equation}
where the bare 3-product $m_3$ is given by
\begin{eqnarray}m_3 \lineup =[ m_2,\mu_2],
\nonumber\\
\lineup = \frac{2}{3}m_2(\xi m_2\otimes\mathbb{I} +\mathbb{I}\otimes \xi m_2),\label{eq:mu32}
\end{eqnarray}
as computed in \cite{WittenSS}.

It is helpful to explain an analogy between the cohomomorphism $\G$ and a classical solution in Berkovits' open superstring field theory \cite{WittenSS}. Note that because $[\n,\M]=0$, the cohomomorphism $\G$ satisfies
\begin{equation}[\n,\G^{-1}[\Q,\G]] = 0.\label{eq:BerkTensor}\end{equation}
This is structurally analogous to the equations of motion of Berkovits' superstring field theory. We can write an equivalent form where $\Q$ and $\n$, and $\G$ and $\G^{-1}$ are interchanged:
\begin{equation}[\Q,\G[\n,\G^{-1}]]=0.\label{eq:BerkTensorint}\end{equation}
Let us verify that $\G$ is a solution. Following \cite{OkWB}, we compute
\begin{eqnarray}
[\n,\G ] \lineup = \int_0^1 dt\, \G(0,t) [\n,\mmu(t)]\G(t,1),\nonumber\\
\lineup = g_o \int_0^1dt\, \G(0,t) \m(t) \G(t,1),\nonumber\\
\lineup = g_o\int_0^1 dt\,\G(0,t)\G(0,t)^{-1}\m_2 \G(0,t)\G(t,1),\nonumber\\
\lineup = g_o \m_2 \int_0^1 dt\, \G(0,1),\nonumber\\
\lineup = g_o \m_2 \G,\label{eq:N0}
\end{eqnarray}
or, equivalently,
\begin{equation}\G\n\G^{-1}= \n-g_o \m_2.\label{eq:N}\end{equation}
Therefore
\begin{equation}[\Q,\G[\n,\G^{-1}]]= -g_o[\Q,\m_2 ] = 0.\end{equation}
This proves that $[\n,\M]=0$, and the products $M_{n+1}$ are in the small Hilbert space. Just like Berkovits' superstring field theory has gauge symmetries, \eq{BerkTensor} admits transformations of $\G$ which relate physically equivalent superstring field theories. Such transformations take the form
\begin{equation}\G \ \to\  \G'=\U \G \V,\label{eq:Gtrans}\end{equation}
where $\U$ and $\V$ are invertible $\Q$- and $\n$- closed cohomomorphisms, respectively:
\begin{equation}[\Q,\U]=0,\ \ \ \ [\n,\V]=0.\end{equation}
Under this transformation the products change as 
\begin{equation}\M \to \M' = \V^{-1}\M \V.\end{equation}
which means that the new theory is related to the old one by a proper field redefinition defined by the cohomomorphism $\V$. Since proper field redefinitions do not effect scattering amplitudes, the transformation from $\G$ to $\G'$ represents a physical equivalence, indeed like a gauge transformation. 

The equations of motion and small Hilbert space constraint for $\PsiA$ can be written
\begin{equation}
\M\frac{1}{1-\PsiA} = 0,\ \ \ \ \ \ \n\frac{1}{1-\PsiA}  = 0.
\end{equation}
Multiplying these equations by $\G$ and making the substitution
\begin{equation}\Psi_A = G^{-1}[\Psi_0],\end{equation}
we obtain
\begin{equation}\Q\frac{1}{1-\Psi_0} = 0,\ \ \ \ (\n-g_o \m_2)\frac{1}{1-\Psi_0}= 0,\end{equation}
where in the second equation we used \eq{N}. Projecting onto the 1-string component of the tensor algebra gives
\begin{eqnarray}Q\Psi_0 \lineup = 0,\label{eq:dualcubic1}\\ 
\eta\Psi_0 - g_o \Psi_0*\Psi_0\lineup =0.\label{eq:dualcubic2}\end{eqnarray}
The equations of motion for $\PsiA$ map into free equations of motion for $\Psi_0$, but the small Hilbert space constraint is not preserved. Instead, we obtain a nonlinear constraint on $\Psi_0$.\footnote{On-shell, $\Psi_0$ can be identified with the string field of the ``dual" cubic theory introduced by Kroyter \cite{Kroyter}. Off-shell, this identification is not quite correct, since for us the equations of motion are $Q\Psi_0 = 0$, while $\eta\Psi_0 -\Psi_0*\Psi_0=0$ is a constraint which is imposed on the string field off-shell. In the dual cubic theory the role of these equations is reversed.} This is the sense that $G[\Psi_A]$ is an improper field redefinition. Note that if we substitute 
\begin{equation}\Psi_0 = \frac{1}{g_o}(\eta e^\Phi)e^{-\Phi}\end{equation}
into \eq{dualcubic1} we obtain the equations of motion of Berkovits superstring field theory,
\begin{equation}Q\Big((\eta e^\Phi)e^{-\Phi}\Big) = 0,\end{equation}
and \eq{dualcubic2} is an identity. Therefore we can map a solution of the Berkovits theory into a solution of the $A_\infty$ theory using
\begin{equation}\PsiA = G^{-1}\left[\frac{1}{g_o}(\eta e^\Phi)e^{-\Phi}\right].\label{eq:BerkAnf}\end{equation}
This does not require fixing a gauge in the Berkovits theory. Only if we want to invert this transformation, mapping a solution of the $A_\infty$ theory into the Berkovits theory, do we need to specify a partial gauge-fixing for the $\eta$ part of the Berkovits gauge invariance. Note that we only mean to define this transformation perturbatively. Nonperturbatively there may be issues with convergence. For example, Berkovits' superstring field theory has a tachyon vacuum solution where $\Psi_0$ lives exclusively in the GSO($+$) sector \cite{Kroyter,cubicvac,BerkVac}. Probably such a solution cannot be transformed into the $A_\infty$ theory using \eq{BerkAnf}, since we would expect the tachyon vacuum in the $A_\infty$ theory to produce an expectation value for the tachyon. See \cite{BerkVac,exotic} for related discussion. 

It is worth mentioning that $\G$ is not the unique cohomomorphism which generates the $A_\infty$ theory from a free theory. We can choose any other $\G'=\U\G\V$ from \eq{Gtrans} such that 
\begin{equation}\M = \V^{-1}\M\V,\end{equation}
which is to say that the field redefinition corresponding to $\V$ is a symmetry of the equations of motion (either a gauge symmetry or global symmetry). For compatibility with the action, we should also require that $\U$ and $\V$ are {\it cyclic} cohomomorphisms (see \eq{Okid}). At first it seems that the BRST closed factor $\U$ can be arbitrary. But for our purposes, there is an additional condition: The map \eq{BerkAnf} from the Berkovits theory into the $A_\infty$ theory must be compatible with the small Hilbert space constraint. This requires that the field redefinition corresponding to $\U$ is a symmetry of the nonlinear constraint \eq{dualcubic2}, which means
\begin{equation}\U(\n - g_o \m_2)\U^{-1} = \n - g_o\m_2.\end{equation}
Subject to these conditions, we can choose a different cohomomorphism $\G'$ to generate the $A_\infty$ theory from a free theory. Similar choices exist for the cohomomorphism $\F$ generating the reduced Berkovits theory from a free theory. This being said, as far as we know our choices of $\G$ and $\F$ are the simplest, and we will stick with them for the remainder of the paper.

\section{Generating the Reduced Berkovits Theory from a Free Theory}
\label{sec:F}

In this section we explain how the reduced Berkovits superstring field theory can be generated by an improper field redefinition of a free theory. The improper field redefinition appears on the right-hand side of the equation
\begin{equation}G[\PsiA]=F[\PsiB],\end{equation}
which relates the $A_\infty$ superstring field theory and the reduced Berkovits superstring field theory. 

The reduced Berkovits theory is defined by the condition \cite{INOT}
\begin{equation}\Phi = \xi\PsiB,\label{eq:Berkgf}\end{equation}
where $\Phi$ is the dynamical field of the Berkovits theory and $\PsiB$ is a degree even, ghost number $1$ and picture number $-1$ NS string field which lives in the small Hilbert space:
\begin{equation}\eta\PsiB = 0.\end{equation}
We will take $\xi$ to be the same operator that appears in the $A_\infty$ theory, though this is not strictly necessary.\footnote{The form of the field redefinition is valid even if we use different $\xi$ operators in the two theories provided that they both satisfy $[\eta,\xi] = g_o$.} In particular we assume that $\xi$ is BPZ even and satisfies
\begin{equation}[\eta,\xi]=g_o.\end{equation}
The action for Berkovits' superstring field theory can be written in many forms. Taking the form of the action written in \cite{heterotic2} with the exponential interpolation of the group element $g(t) = e^{t\Phi}$, the reduced Berkovits action can be written 
\begin{equation}S_B=-\frac{1}{g_o^2}\int_0^1 dt\,\left\langle \xi\PsiB,Q\Big((\eta e^{t\xi\PsiB}) e^{-t\xi\PsiB}\Big)\right\rangle_L .\label{eq:BerkS}\end{equation}
Note that the action is finite in the $g_o\to 0$ limit since the $1/g_o^2$ is compensated by the vanishing of $\eta$ and the large Hilbert space BPZ inner product. The equations of motion can be expressed in the form
\begin{equation}Q\Big((\eta e^{\xi\PsiB})e^{-\xi\PsiB}\Big)= 0.\label{eq:BEOM2}\end{equation}
Therefore it is natural to suppose that the improper field redefinition defining the reduced Berkovits theory is
\begin{equation}
F[\PsiB]  = \frac{1}{g_o}(\eta e^{\xi \PsiB})e^{-\xi\PsiB},
\label{eq:Fexp}\end{equation}
or, acting $\eta$ in the above expression,
\begin{equation}F[\PsiB]= \int_0^1 dt\, e^{t\xi\PsiB}\,\PsiB\, e^{-t\xi\PsiB}.\end{equation}
Note that this is well defined in the $g_o\to 0$ limit. 

\subsection{Action in Small Hilbert Space}
\label{subsec:smallS}

To compare to the $A_\infty$ superstring field theory, it is helpful to express the reduced Berkovits action in small Hilbert space, as follows:
\begin{equation}S_B=\frac{1}{2}\omega_S(\PsiB,Q\PsiB) + \frac{1}{3}\omega_S(\PsiB,\tilde{B}_2(\PsiB,\PsiB)) + \frac{1}{4}\omega_S(\PsiB,\tilde{B}_3(\PsiB,\PsiB,\PsiB))+...\ .\label{eq:SBsmall}\end{equation}
Here $Q,\tilde{B}_2,\tilde{B}_3,...$ are a sequence of degree odd cyclic multi-string products which live in the small Hilbert space:
\begin{equation}[\eta,\tilde{B}_{n+1}]=0.\end{equation}
Let us compute the action in this form out to cubic order:
\begin{eqnarray}S_B \lineup= -\frac{1}{g_o}\int_0^1 dt\, \left\langle \xi\PsiB,\,tQ\PsiB + \frac{t^2}{2} Q[\xi\PsiB,\PsiB]+...\right\rangle_L,\nonumber\\
\lineup = -\frac{1}{g_o}\left[\frac{1}{2}\big\langle \xi\PsiB,Q\PsiB\big\rangle_L + \frac{1}{6}\big\langle \xi\PsiB,Q[\xi\PsiB,\PsiB]\big\rangle_L +...\right],\nonumber\\
\lineup = -\frac{1}{g_o}\left[\frac{1}{2}\big\langle \xi\PsiB,Q\PsiB\big\rangle_L + \frac{1}{6}\big\langle \xi\PsiB,\eta \xi Q[\xi\PsiB,\PsiB]\big\rangle_L +...\right],
\nonumber\\
\lineup = \frac{1}{2}\big\langle \PsiB,Q\PsiB\big\rangle_S + \frac{1}{6}\big\langle \PsiB,\eta \xi Q[\xi\PsiB,\PsiB]\big\rangle_S +...\ ,
\nonumber\\
\lineup = \frac{1}{2}\big\langle \PsiB,Q\PsiB\big\rangle_S + \frac{1}{6}\Big\langle \PsiB,\Big( Q[\xi\PsiB,\PsiB]+2\xi Q(\PsiB^2)\Big)\Big\rangle_S +...\ .
\end{eqnarray}
Here we define the commutator of string fields, 
\begin{equation}[A,B] \equiv A*B-(-1)^{\eps(A)}B*A.\end{equation} 
which is graded with respect to Grassmann parity. In the first step we expanded \eq{BerkS} out to the cubic vertex; in the second step we integrated over $t$; in the third step we inserted a trivial factor 
$\eta\xi$ in the cubic vertex to project the second entry of the BPZ inner product into the small Hilbert space; and in the fourth step we replaced the large Hilbert space BPZ inner product with the small Hilbert space BPZ inner product using
\begin{equation}\langle \xi A,B\rangle_L = -g_o \langle A,B\rangle_S,\label{eq:BPZls}\end{equation}
for $A,B$ in the small Hilbert space; in the final step we acted $\eta$ on the second entry of the BPZ inner product in the cubic vertex. Inserting the appropriate signs, we can translate this expression to the degree grading:
\begin{equation}
S_B = \frac{1}{2}\omega_S(\PsiB,Q\PsiB) + \frac{1}{3}\omega_S\left(\PsiB,-\frac{1}{2}\Big(Qm_2(\xi\PsiB,\PsiB)+Qm_2(\PsiB,\xi\PsiB)-2\xi Q m_2(\PsiB,\PsiB)\Big)\right)+...\ .
\end{equation}
This is almost what we want, but the multi-string product in the cubic vertex,
\begin{equation}\Bnc_2=-\frac{1}{2}\Big(Q m_2(\xi\otimes\mathbb{I}+\mathbb{I}\otimes\xi)-2\xi Qm_2\Big),\end{equation}
is not cyclic. To get $\tilde{B}_2$ we should take the cyclic projection:
\begin{equation}\tilde{B}_2 = \frac{1}{3}\Big(1+\omega +\omega^2\Big)\circ\Bnc_2.\end{equation}
Computing we find
\begin{equation}
\tilde{B}_2 = -\frac{1}{2}\Big(Q m_2(\xi\otimes \mathbb{I}+\mathbb{I}\otimes\xi) -\xi m_2(Q\otimes\mathbb{I}+\mathbb{I}\otimes Q) +m_2(\xi Q\otimes\mathbb{I}+\mathbb{I}\otimes\xi Q)\Big).
\end{equation}
Acting on $\PsiB\otimes\PsiB$ and reverting to the Grassmann grading gives
\begin{equation}\tilde{B}_2(\PsiB,\PsiB)=\frac{1}{2}\Big(Q[\xi\PsiB,\PsiB] - \xi[Q\PsiB,\PsiB]+[\xi Q\PsiB,\PsiB]\Big).\label{eq:tB2}\end{equation}
For comparison, to the 2-product of the $A_\infty$ theory is
\begin{equation}M_2(\PsiA,\PsiA)=\frac{1}{3}\Big(X\PsiA^2 + [X\PsiA,\PsiA]\Big).\end{equation}
We can check that $\tilde{B}_2$ is in the small Hilbert space
\begin{equation}[\eta,\tilde{B}_2]= - Qm_2 -m_2(Q\otimes\mathbb{I}+\mathbb{I}\otimes Q) = 0.\end{equation}
However, $Q$ is not a derivation of $\tilde{B}_2$:
\begin{equation}[Q,\tilde{B}_2]\neq 0.\end{equation}
This means that the cyclic products $Q,\tilde{B}_2,\tilde{B}_3,...$ {\it do not} form an $A_\infty$ algebra. Perhaps this is surprising, since in section \ref{subsec:FR} we argued that field redefinitions always map $A_\infty$ algebras into $A_\infty$ algebras. The resolution to this puzzle is that the reduced Berkovits theory {\it does} have a natural $A_\infty$ structure, but it is not cyclic.  Therefore the $A_\infty$ structure cannot be realized in the action.

Let us demonstrate---up to cubic order and when $g_o=0$---that the reduced Berkovits superstring field theory can be obtained by field redefinition from a free theory. Take the free action
\begin{equation}S_\mathrm{free} = \frac{1}{2}\omega_S(\Psi_0,Q\Psi_0),\end{equation}
and plug in the field redefinition 
\begin{equation}\Psi_0 = \PsiB + F_2(\PsiB,\PsiB) + \mathrm{higher\ orders},\label{eq:FR3}\end{equation}
where $F_2$ is a degree even 2-product of the form
\begin{equation}F_2= -\frac{1}{2}m_2(\xi\otimes \mathbb{I}+ \mathbb{I}\otimes \xi).\label{eq:mgu2}\end{equation}
Here we must take $g_o=0$ otherwise the field redefinition cannot be defined in the small Hilbert space. The action becomes
\begin{eqnarray}
S_\mathrm{free}\lineup = \frac{1}{2}\omega_S(\PsiB,Q\PsiB)+\frac{1}{2}\omega_S(\PsiB,QF_2(\PsiB,\PsiB)) + \frac{1}{2}\omega_S(F_2(\PsiB,\PsiB),Q\PsiB)+\mathrm{higher\ orders},\nonumber\\
\lineup = \frac{1}{2}\omega_S(\PsiB,Q\PsiB)+\omega_S(\PsiB,Q F_2(\PsiB,\PsiB)) +\mathrm{higher\ orders},\nonumber\\
\lineup = \frac{1}{2}\omega_S(\PsiB,Q\PsiB)+\frac{1}{3}\omega_S\Big(\PsiB,QF_2(\PsiB,\PsiB) - (\omega\circ F_2)(Q\PsiB,\PsiB) - (\omega^2\circ F_2)(\PsiB,Q\PsiB)\Big) 
\nonumber\\
\lineup\ \ \ +\mathrm{higher\ orders},
\end{eqnarray}
where in the third step we took the cyclic projection of $3\cdot QF_2$ to obtain the cyclic product 
\begin{equation}QF_2 -(\omega \circ F_2)(Q\otimes\mathbb{I}) - (\omega^2 \circ F_2)(\mathbb{I}\otimes Q).\label{eq:QFcyc}\end{equation}
Compute $\omega\circ F_2$:
\begin{eqnarray}
\langle \omega_S| \mathbb{I}\otimes F_2 \lineup = -\frac{1}{2}\langle \omega_S|\mathbb{I}\otimes m_2(\xi\otimes \mathbb{I}+\mathbb{I}\otimes \xi),\nonumber\\
\lineup = -\frac{1}{2}\langle \omega_S| (-m_2(\mathbb{I}\otimes \xi)\otimes \mathbb{I} -m_2\otimes \xi),\nonumber\\
\lineup = -\frac{1}{2}\langle \omega_S| (m_2(-\mathbb{I}\otimes \xi)\otimes \mathbb{I} +\xi m_2\otimes \mathbb{I}),\nonumber\\
\lineup = -\langle \omega_S|(\omega\circ F_2)\otimes \mathbb{I}.
\end{eqnarray}
Therefore 
\begin{equation}\omega\circ F_2 = \frac{1}{2}\Big(\xi m_2 -m_2(\mathbb{I}\otimes \xi)\Big).\end{equation}
A similar computation gives
\begin{equation}\omega^2\circ F_2 = \frac{1}{2}\Big(\xi m_2 - m_2(\xi\otimes\mathbb{I})\Big).\end{equation}
Plugging in to \eq{QFcyc} we obtain
\begin{eqnarray}\lineup QF_2 -(\omega \circ F_2)(Q\otimes\mathbb{I}) - (\omega^2 \circ F_2)(\mathbb{I}\otimes Q) = \nonumber\\
\lineup \ \ \ \ \ \ \ \ \ \ \ \ \ \ \ \ \ \ \ 
-\frac{1}{2}\Big(Q m_2(\xi\otimes \mathbb{I}+\mathbb{I}\otimes\xi) -\xi m_2(Q\otimes\mathbb{I}+\mathbb{I}\otimes Q) +m_2(\xi Q\otimes\mathbb{I}+\mathbb{I}\otimes\xi Q)\Big).
\end{eqnarray}
Comparing with \eq{tB2} we therefore have 
\begin{equation}\tilde{B}_2 = QF_2 -(\omega \circ F_2)(Q\otimes\mathbb{I}) - (\omega^2 \circ F_2)(\mathbb{I}\otimes Q),\end{equation}
and the action can be written
\begin{equation} S_\mathrm{free} = \frac{1}{2}\omega_S(\PsiB,Q\PsiB) + \frac{1}{3}\omega_S(\PsiB,\tilde{B}_2(\PsiB,\PsiB))+\mathrm{higher\ orders}.\end{equation}
Therefore---at least to this order and when $\xi$ is in the small Hilbert space---the action for the reduced Berkovits superstring field theory is generated from a free action by field redefinition. Note that, while $\tilde{B}_2$ can be written in terms of the BRST operator and a ``gauge product" $F_2$, it is not in the form of a BRST commutator. This is why $[Q,\tilde{B}_2]\neq 0$, and the cyclic products do not realize $A_\infty$.

The main part of our analysis does not require the explicit form of the higher cyclic products $\tilde{B}_3,\tilde{B}_4$ and so on. However, for completeness we give a derivation of the general formula in appendix \ref{app:Bcyc}.

\subsection{Non-cyclic $A_\infty$ Structure}

The reduced Berkovits theory has a natural $A_\infty$ structure defined by:
\begin{equation}\B=\F^{-1}\Q\F,\end{equation}
where $\F$ is the cohomomorphism corresponding to $F[\PsiB]$. In particular, we can express the Berkovits equations of motion
\begin{equation}\B\frac{1}{1-\PsiB}=0,\end{equation}
or equivalently
\begin{equation}0= Q\PsiB + B_2(\PsiB,\PsiB) + B_3(\PsiB,\PsiB,\PsiB) + ... \ ,\end{equation}
where $Q,B_2,B_3,...$ are multi-string products of odd degree which satisfy the relations of an $A_\infty$ algebra.

To construct the cohomomorphism $\F$ we need a sequence of multi-string products $F_0,F_1,F_2,...$ such that the equation 
\begin{equation}F[\PsiB]=\pi_1\F\frac{1}{1-\PsiB}\end{equation}
is obeyed. Expanding $F[\PsiB]$,
\begin{equation}
F[\PsiB] =\PsiB + \frac{1}{2!}[\xi\PsiB,\PsiB] + \frac{1}{3!}[\xi\PsiB,[\xi\PsiB,\PsiB]] +...\ ,
\end{equation}
the products of the cohomomorphism must be defined to satisfy
\begin{eqnarray}
F_0 \lineup = 0,\\
F_1(\PsiB)\lineup = \PsiB,\\
F_2(\PsiB,\PsiB)\lineup = \frac{1}{2!}[\xi\PsiB,\PsiB],\\
F_3(\PsiB,\PsiB,\PsiB) \lineup = \frac{1}{3!}[\xi\PsiB,[\xi\PsiB,\PsiB]],\\
\lineup \vdots \nonumber \ .
\end{eqnarray}
Factoring out tensor products of $\PsiB$, and switching from the Grassmann grading to degree, suggests a natural definition for the products: 
\begin{eqnarray}
F_0\lineup \equiv 0,\\
F_1\lineup \equiv \mathbb{I},\\
F_2\lineup \equiv -\frac{1}{2!}m_2(\xi\otimes\mathbb{I} +\mathbb{I}\otimes\xi),\label{eq:F2}\\
F_3\lineup \equiv \frac{1}{3!}m_2(\xi\otimes m_2(\xi\otimes\mathbb{I}+\mathbb{I}\otimes\xi) +m_2(\xi\otimes\mathbb{I}+\mathbb{I}\otimes\xi)\otimes\xi),\label{eq:F3}\\
\lineup \vdots \nonumber\ .
\end{eqnarray}
The products can be expressed recursively:
\begin{equation}F_{n+2}=-\frac{1}{n+2}m_2(\xi\otimes F_{n+1} + F_{n+1}\otimes\xi).\label{eq:Frec}\end{equation}
Plugging into \eq{H}  gives an expression for $\F$
\begin{equation}\F = \mathbb{I}_{T\mathcal{H}} + \sum_{\ell=1}^{\infty} \sum_{{k_1,...,k_{\ell}\geq 0 \atop k_1+...+k_{\ell} \geq 1}} (F_{k_1+1}\otimes ...\otimes F_{k_{\ell}+1})\pi_{k_1+...+k_{\ell}+\ell}.\end{equation}
Here we used $F_0=0$ and $F_1=\mathbb{I}$ to pull an identity operator out from the multiple sum. This allows us to compute $\F^{-1}$ as a geometric series, finding a sequence of products:
\begin{eqnarray}
F_0^{-1}\lineup = 0,\\
F_1^{-1}\lineup = \mathbb{I},\\
F_2^{-1}\lineup = -F_2,\\
F_3^{-1}\lineup = -F_3 +F_2(F_2\otimes\mathbb{I}+\mathbb{I}\otimes F_2),\\
F_4^{-1}\lineup = -F_4 +F_2(F_3\otimes\mathbb{I}+\mathbb{I}\otimes F_3) 
+ F_3(F_2\otimes\mathbb{I}\otimes\mathbb{I}+\mathbb{I}\otimes F_2\otimes\mathbb{I}+\mathbb{I}\otimes\mathbb{I}\otimes F_2)\nonumber\\
\lineup \ \ \ -F_2(F_2\otimes F_2) - F_2(F_2(F_2\otimes\mathbb{I}+\mathbb{I}\otimes F_2)\otimes\mathbb{I}+\mathbb{I}\otimes F_2(F_2\otimes\mathbb{I}+\mathbb{I}\otimes F_2)),\\
\lineup \vdots \nonumber\ .
\end{eqnarray}
Computing $\F^{-1}\Q\F$ gives expressions for the $2$- and $3$-string products:
\begin{eqnarray}
B_2 \lineup = \frac{1}{2}m_2(X\otimes\mathbb{I}+\mathbb{I}\otimes X),\\
B_3 \lineup = -\frac{1}{6}m_2(X\otimes m_2(\xi\otimes\mathbb{I} + \mathbb{I}\otimes\xi)+m_2(\xi\otimes\mathbb{I} + \mathbb{I}\otimes\xi)\otimes X)\nonumber\\
\lineup\ \ \ +\frac{1}{12}m_2(\xi\otimes m_2(X\otimes\mathbb{I} + \mathbb{I}\otimes X)-m_2(X\otimes\mathbb{I} + \mathbb{I}\otimes X)\otimes \xi)
\nonumber\\
\lineup \ \ \ +\frac{1}{4}m_2(\mathbb{I}\otimes \xi m_2(X\otimes\mathbb{I} + \mathbb{I}\otimes X)+\xi m_2(X\otimes\mathbb{I} + \mathbb{I}\otimes X)\otimes \mathbb{I}).\label{eq:B3}
\end{eqnarray}
By inspection we can see  
\begin{equation}[Q,B_2] = 0,\end{equation}
since both $m_2$ and $X$ are BRST closed. Computing $[Q,B_3]$ amounts to replacing $\xi$ in \eq{B3} with $X$, with the appropriate sign for anticommutation from the BRST operator:
\begin{eqnarray}
[Q,B_3] \lineup = \left(-\frac{1}{6}-\frac{1}{12}\right)m_2(X\otimes m_2(X\otimes\mathbb{I} + \mathbb{I}\otimes X)-m_2(X\otimes\mathbb{I} + \mathbb{I}\otimes X)\otimes X)
\nonumber\\
\lineup\ \ \ -\frac{1}{4}m_2(\mathbb{I}\otimes X m_2(X\otimes\mathbb{I} + \mathbb{I}\otimes X)+X m_2(X\otimes\mathbb{I} + \mathbb{I}\otimes X)\otimes\mathbb{I}),
\nonumber\\
 \lineup = -\frac{1}{4}m_2(X\otimes m_2(X\otimes\mathbb{I} + \mathbb{I}\otimes X)-m_2(X\otimes\mathbb{I} + \mathbb{I}\otimes X)\otimes X)
\nonumber\\
\lineup\ \ \ -\frac{1}{4}m_2(\mathbb{I}\otimes X m_2(X\otimes\mathbb{I} + \mathbb{I}\otimes X)+X m_2(X\otimes\mathbb{I} + \mathbb{I}\otimes X)\otimes\mathbb{I}),\nonumber\\
\lineup = -B_2(B_2\otimes\mathbb{I}+\mathbb{I}\otimes B_2),\nonumber\\
\lineup = -\frac{1}{2}[B_2,B_2].
\end{eqnarray}
This confirms the $A_\infty$ relations up to cubic order. One can also check that $B_2$ and $B_3$ are in the small Hilbert space. However,
since
\begin{equation}\omega\circ B_2 = \frac{1}{2}(m_2(\mathbb{I}\otimes X) +X m_2)\neq B_2.\end{equation}
the products do not define a cyclic $A_\infty$ algebra. 

Acting on $\PsiB$ and reverting to the Grassmann grading, the 2- and 3-string products take the form
\begin{eqnarray}
B_2(\PsiB,\PsiB)\lineup = \frac{1}{2}[X\PsiB,\PsiB],\label{eq:B2G}\\
B_3(\PsiB,\PsiB,\PsiB)\lineup = \frac{1}{6}[X\PsiB,[\xi\PsiB,\PsiB]]-\frac{1}{12}[\xi\PsiB,[X\PsiB,\PsiB]]+\frac{1}{4}[\PsiB,\xi[X\PsiB,\PsiB]].\ \ \ \ \ \label{eq:B3G}
\end{eqnarray}
It is instructive to see how these products are derived from the Berkovits equations of motion as they are usually expressed. Expanding the equations of motion up to cubic order gives
\begin{eqnarray}
0\lineup = Q\left(\int_0^1 dt\, e^{\xi\PsiB}\PsiB e^{-\xi\PsiB}\right),\nonumber\\
\lineup = Q\left(\PsiB +\frac{1}{2!}[\xi\PsiB,\PsiB] + \frac{1}{3!}[\xi\PsiB,[\xi\PsiB,\PsiB]]\right)+\mathrm{higher\ orders},\nonumber\\
\lineup = Q\PsiB +\frac{1}{2!}[X\PsiB,\PsiB] + \frac{1}{3!}[X\PsiB,[\xi\PsiB,\PsiB]]+\frac{1}{3!}[\xi\PsiB,[X\PsiB,\PsiB]],\nonumber\\
\lineup\ \ \ -\frac{1}{2!}[\xi Q\PsiB,\PsiB]+\frac{1}{2!}[\xi\PsiB,Q\PsiB]-\frac{1}{3!}[\xi Q\PsiB,[\xi\PsiB,\PsiB]]-\frac{1}{3!}[\xi\PsiB,[\xi Q\PsiB,\PsiB]]\nonumber\\
\lineup\ \ \ +\frac{1}{3!}[\xi\PsiB,[\xi\PsiB,Q\PsiB]]
+ \mathrm{higher\ orders}.
\end{eqnarray}
Note that we can eliminate $Q\PsiB$ from nonlinear terms in this equation, since $Q\PsiB$ is already determined by the equations of motion at lower orders. In the quadratic terms we can substitute
\begin{equation}
Q\PsiB = - \frac{1}{2!}[X\PsiB,\PsiB] +\frac{1}{2!}[\xi Q\PsiB,\PsiB]-\frac{1}{2!}[\xi\PsiB,Q\PsiB]+\mathrm{higher\ orders},
\end{equation}
and in the cubic terms
\begin{equation}Q\PsiB = 0 +\mathrm{higher\ orders}.\label{eq:cubel}\end{equation}
Plugging in,
\begin{eqnarray}
0\lineup = Q\PsiB +\frac{1}{2!}[X\PsiB,\PsiB] + \frac{1}{3!}[X\PsiB,[\xi\PsiB,\PsiB]]+\frac{1}{3!}[\xi\PsiB,[X\PsiB,\PsiB]]\nonumber\\
\lineup\ \ \ -\frac{1}{2!}\left[\xi \left(- \frac{1}{2!}[X\PsiB,\PsiB] +\frac{1}{2!}[\xi Q\PsiB,\PsiB]-\frac{1}{2!}[\xi\PsiB,Q\PsiB]\right),\PsiB\right]\nonumber\\
\lineup\ \ \ +\frac{1}{2!}\left[\xi\PsiB,\left(- \frac{1}{2!}[X\PsiB,\PsiB] +\frac{1}{2!}[\xi Q\PsiB,\PsiB]-\frac{1}{2!}[\xi\PsiB,Q\PsiB]\right)\right] + \mathrm{higher\ orders},\nonumber\\
\lineup = Q\PsiB +\frac{1}{2!}[X\PsiB,\PsiB] + \frac{1}{3!}[X\PsiB,[\xi\PsiB,\PsiB]]+\frac{1}{3!}[\xi\PsiB,[X\PsiB,\PsiB]]\nonumber\\
\lineup\ \ \ +\frac{1}{(2!)^2}\left[\xi [X\PsiB,\PsiB],\PsiB\right]-\frac{1}{(2!)^2}\left[\xi\PsiB,[X\PsiB,\PsiB]\right] + \mathrm{higher\ orders}.
\end{eqnarray}
Collecting terms,
\begin{eqnarray}
0\lineup = Q\PsiB +\frac{1}{2!}[X\PsiB,\PsiB] +\frac{1}{6}[X\PsiB,[\xi\PsiB,\PsiB]]
-\frac{1}{12}[\xi\PsiB,[X\PsiB,\PsiB]]+\frac{1}{4}[\PsiB,\xi[X\PsiB,\PsiB]]\nonumber\\
\lineup\ \ \ +\mathrm{higher\ orders},\nonumber\\
\lineup = Q\PsiB + B_2(\PsiB,\PsiB) + B_3(\PsiB,\PsiB,\PsiB) + \mathrm{higher\ orders}.
\end{eqnarray}
In this way we recover the $A_\infty$ products of the Berkovits theory. 

In appendix \ref{app:nF} we show 
\begin{equation}\F\n\F^{-1} = \n - g_o\m_2.\label{eq:Fn}\end{equation}
This means that the cohomomorphism $\F$ satisfies the same equation \eq{BerkTensor} as $\G$,
\begin{equation}[\n,\F^{-1}[\Q,\F]]=0.\end{equation}
Therefore the products $B_n$ are in the small Hilbert space. We can develop the analogy with the $A_\infty$ theory further. Define the cohomomorphism 
\begin{equation}\F(0,t) = \mathbb{I}_{T\mathcal{H}} + \sum_{\ell=1}^{\infty} \sum_{{k_1,...,k_{\ell}\geq 0 \atop k_1+...+k_{\ell}\geq 1}} 
t^{k_1+...+k_{\ell} }(F_{k_1+1}\otimes ...\otimes F_{k_{\ell}+1})\pi_{k_1+...+k_{\ell}+\ell}\end{equation}
such that $\F(0,1)=\F$. This is the cohomomorphism obtained upon replacing $F_{n+1}\to t^n F_{n+1}$. Following the argument of appendix \ref{app:nF}, we can show that $\F(0,t)$ satisfies
\begin{equation}\F(0,t)\n\F(0,t)^{-1} = \n - t g_o \m_2.\end{equation}
Next, define generating functions for products, gauge products, and bare products analogously to \eq{mmuG}-\eq{mG} of the $A_\infty$ theory
\begin{eqnarray}
\B(t) \lineup = \F(0,t)^{-1}\Q\F(0,t),\\
\bbeta(t) \lineup =\F(0,t)^{-1}\frac{d}{dt}\F(0,t),\\
\b(t) \lineup = \F(0,t)^{-1}\m_2\F(0,t).
\end{eqnarray}
Compute
\begin{eqnarray}
[\n,\bbeta(t)]\lineup =\left[\F(0,t)^{-1}\Big(\n - tg_o\m_2\Big)\F(0,t),\,\F(0,t)^{-1}\frac{d}{dt}\F(0,t)\right],\nonumber\\
\lineup = \F(0,t)^{-1}\Big(\n - tg_o\m_2\Big)\frac{d}{dt}\F(0,t) -\F(0,t)^{-1}\left(\frac{d}{dt}\F(0,t)\right)\F(0,t)^{-1}\Big(\n - tg_o\m_2\Big)\F(0,t),
\nonumber\\
\lineup = \frac{d}{dt}\Big(\F(0,t)^{-1}(\n - tg_o\m_2)\F(0,t)\Big)-\F(0,t)^{-1}\left(\frac{d}{dt}(\n - tg_o\m_2)\right)\F(0,t),
\nonumber\\
\lineup = \frac{d}{dt}\n +g_o \F(0,t)^{-1}\m_2\F(0,t),\nonumber\\
\lineup = g_o \b(t).
\end{eqnarray}
Therefore the generating functions satisfy the equations
\begin{eqnarray}
\frac{d}{dt}\B(t)\lineup = [\B(t),\bbeta(t)],\label{eq:Brec1}\\
\frac{d}{dt}\b(t)\lineup = [\b(t),\bbeta(t)],\\
\ [\n,\bbeta(t)] \lineup = g_o \b(t).\label{eq:Brec2}
\end{eqnarray}
This is precisely analogous to equations \eq{rec1}-\eq{rec2} of the $A_\infty$ theory. What ultimately distinguishes the two theories is the choice of solution 
to \eq{Brec2} which expresses the gauge products $\beta_{n+2}$ in terms of the bare products $b_{n+2}$. The solution which defines the $A_\infty$ theory is \eq{rec2inv}, while the solution which defines the reduced Berkovits theory is implicitly defined by $\F$, which has been {\it a priori} provided.

With a little thought, it is clear that the structural similarity between the reduced Berkovits theory and the $A_\infty$ theory has nothing to do with the precise definition of the products in the respective theories. Any superstring field theory in the small Hilbert space which is related to the $A_\infty$ theory by field redefinition
\begin{equation}\PsiA = H[\Psi]\end{equation}
can be constructed from ``free theory" by the field redefinition
\begin{equation}\Psi_0 = G^{-1}[H[\Psi]].\end{equation}
Turning this into a cohomomorphism, we conclude that the generic NS open superstring field theory has an $A_\infty$ structure at least at the level of the equations of motion, and that there is a corresponding set of products, gauge products, and bare products which satisfy the analogue of \eq{Brec1}-\eq{Brec2}.

\section{Equivalence of Actions}
\label{sec:equivalence}

In summary, the field redefinition from the $A_\infty$ theory to the reduced Berkovits theory is defined by the cohomomorphism
\begin{equation}\E = \G^{-1}\F.\label{eq:GF}\end{equation}
In particular this implies 
\begin{equation}\PsiA = E[\PsiB],\label{eq:Eexp}\end{equation}
where $\PsiA$ is the field of the $A_\infty$ theory, $\PsiB$ is the field of the reduced Berkovits theory, and the field redefinition is
\begin{eqnarray}E[\PsiB] \lineup = \pi_1\E\frac{1}{1-\PsiB}, \nonumber\\
\lineup = \pi_1\G^{-1}\F\frac{1}{1-\PsiB},\nonumber\\
\lineup = \pi_1\G^{-1} \frac{1}{1-F[\PsiB]}, \nonumber\\
\lineup = \pi_1 \frac{1}{1-G^{-1}[F[\PsiB]]}. \nonumber\\
\lineup = G^{-1}[F[\PsiB]]
\end{eqnarray}
The cohomomorphism $\E$ maps between the $A_\infty$ structures,
\begin{equation}\E \B = \M \E,\end{equation}
and lives in the small Hilbert space
\begin{equation}[\n,\E] = 0.\end{equation}
Computing \eq{GF}, we find that $\E$ corresponds to a sequence of products 
\begin{eqnarray}
E_0 \lineup = 0,\label{eq:E0}\\
E_1 \lineup = \mathbb{I},\\
E_2 \lineup = F_2 - \mu_2,\label{eq:E2}\\
E_3\lineup = F_3 - \mu_2(F_2\otimes \mathbb{I}+\mathbb{I}\otimes F_2)-\frac{1}{2}\mu_3 + \frac{1}{2}\mu_2(\mu_2\otimes\mathbb{I}+\mathbb{I}\otimes\mu_2),\label{eq:E3}\\
\lineup \vdots \ , \nonumber
\end{eqnarray}
so that the field redefinition is explicitly (up to cubic order)
\begin{eqnarray}\PsiA \lineup = \PsiB + F_2(\PsiB,\PsiB) - \mu_2(\PsiB,\PsiB)\nonumber\\
\lineup\ \ \ +F_3(\PsiB,\PsiB,\PsiB) -\mu_2(F_2(\PsiB,\PsiB),\PsiB)-\mu_2(\PsiB,F_2(\PsiB,\PsiB))\nonumber\\
\lineup\ \ \ -\frac{1}{2}\mu_3(\PsiB,\PsiB,\PsiB)+\frac{1}{2}\mu_2(\mu_2(\PsiB,\PsiB),\PsiB)+\frac{1}{2}\mu_2(\PsiB,\mu_2(\PsiB,\PsiB))\nonumber\\
\lineup\ \ \ +... \ .
\end{eqnarray}
For reference, the products $F_2$ and $F_3$ are described in \eq{F2} and \eq{F3}, and the products $\mu_2$ and $\mu_3$ are described in equations
\eq{mu2}, \eq{mu3}, and \eq{mu32}.

Our discussion so far establishes equivalence of the theories at the level of the equations of motion (at least perturbatively). But ideally we would like to establish equivalence of the actions. That is, we should have 
\begin{equation}S_B[\PsiB]=S_A[E[\PsiB]],\label{eq:SBA}\end{equation}
where $S_B$ is the action of the reduced Berkovits theory and $S_A$ is the action of the $A_\infty$ theory. The equality of the actions is equivalent to the equality of the cyclic products which define the vertices:\footnote{Equality of vertices only implies equality of the cyclic products up to terms which are antisymmetric upon interchange of two or more inputs. However, such terms cannot be constructed out of the ingredients $Q,m_2,\xi$ which define the products of the $A_\infty$ theory and the partially gauge fixed Berkovits theory. We would need, for example, an additional operation which maps $A\otimes B$ into $B\otimes A$.} 
\begin{eqnarray}
\tilde{B}_{n+1} = -\sum_{m=0}^{\infty} \sum_{{k_0,...,k_{m+1}\geq 0 \atop k_0+...+k_{m+1} = n-m}}\sum_{j=0}^{k_0} (\omega^{j+1}\circ E_{k_0+1})(\mathbb{I}^{\otimes j}\otimes M_{m+1}(E_{k_1+1}\otimes...\otimes E_{k_{m+1}+1})\otimes\mathbb{I}^{\otimes k_0-j}),\ \ \ \ 
\label{eq:tBM}\end{eqnarray}
where the right hand side is derived by expanding the field redefinition $E[\PsiB]$ into component products inside the $A_\infty$ action, collecting terms into $(n+2)$-string vertices and extracting the corresponding $(n+1)$-string cyclic product. Let's see how this formula works for the 2-product. We should have
\begin{equation}\tilde{B}_2 = -(\omega\circ E_1)Q E_2 - (\omega\circ E_2)(Q E_1\otimes \mathbb{I}) - (\omega^2\circ E_2)(\mathbb{I}\otimes QE_1)
-(\omega\circ E_1)M_2(E_1\otimes E_1).
\end{equation}
Noting $E_1=\mathbb{I}$ and $\omega\circ \mathbb{I} = -\mathbb{I}$, this simplifies to
\begin{equation}\tilde{B}_2 = Q E_2 - (\omega\circ E_2)(Q\otimes \mathbb{I}) - (\omega^2\circ E_2)(\mathbb{I}\otimes Q)+M_2.
\end{equation}
Plugging in $E_2$ from \eq{E2} gives
\begin{equation}\tilde{B}_2 = Q (F_2 - \mu_2) - (\omega\circ F_2 - \mu_2)(Q\otimes \mathbb{I}) - (\omega^2\circ F_2 -\mu_2)(\mathbb{I}\otimes Q)+M_2,
\end{equation}
where we use the fact that $\mu_2$ is cyclic. Thus
\begin{eqnarray}\tilde{B}_2 \lineup = Q F_2 - (\omega\circ F_2)(Q\otimes \mathbb{I}) - (\omega^2\circ F_2)(\mathbb{I}\otimes Q)- Q\mu_2+\mu_2(Q\otimes\mathbb{I}+\mathbb{I}\otimes Q) +M_2,\nonumber\\
\lineup = Q F_2 - (\omega\circ F_2)(Q\otimes \mathbb{I}) - (\omega^2\circ F_2)(\mathbb{I}\otimes Q).
\end{eqnarray}
which is the correct expression for $\tilde{B}_2$. The corresponding calculation for higher products is an extension of the computation given in appendix \ref{app:Bcyc}, but is rather unwieldy. 

One way to prove the equivalence of the actions is to formulate the $A_\infty$ theory in WZW-like form, as discussed in \cite{OkWB}. Here we will take a different approach. The proof goes as follows. First we show that the $A_\infty$ action can be written as a free action
\begin{equation}S_A[\PsiA] = \frac{1}{2}\omega_S(G[\PsiA],QG[\PsiA])\label{eq:SAfree}\end{equation}
when $\xi$ is in the small Hilbert space. Then we show that the reduced Berkovits action can be written as a free action
\begin{equation}S_B[\PsiB] = \frac{1}{2}\omega_S(F[\PsiB],QF[\PsiB])\label{eq:SBfree}\end{equation}
when $\xi$ is in the small Hilbert space. It follows that if we take
\begin{equation}G[\PsiA] = F[\PsiB],\label{eq:idproof}\end{equation}
the actions are equal when $\xi$ is in the small Hilbert space. This argument is also implies that the actions are equal when $\xi$ is in the large Hilbert space, for the following reason. If we are able to demonstrate \eq{SAfree} and \eq{SBfree}, we can conclude that \eq{tBM} holds when $\xi$ is in the small Hilbert space. However, the computation of \eq{tBM} is a strictly algebraic procedure which only requires knowledge that $\xi$ is BPZ even, and otherwise is completely blind to the concrete definition of $\xi$. Therefore, if the computation of \eq{tBM} works when $\xi$ is in the small Hilbert space, the same computation will also work when $\xi$ is in the large Hilbert space. Therefore the $A_\infty$ and reduced Berkovits actions would be related by the field redefinition \eq{GF} even if the coupling constant is nonzero. With this in mind, we now turn to the proof that $G[\PsiA]$ and $F[\PsiB]$ map the respective nonlinear actions to a free action when $\xi$ is in the small Hilbert space. 

\subsection{Mapping the $A_\infty$ Action into a Free Action}
\label{subsec:Afree}

We want to show that the $A_\infty$ action,
\begin{equation}S_A = \sum_{n=0}^\infty \frac{1}{n+2}\omega_S(\PsiA,M_{n+1}(\PsiA,...,\PsiA)),\end{equation}
can be mapped to a free action when $\xi$ is in the small Hilbert space. To do this, it is helpful to introduce an auxiliary parameter $t$ and write the action in the form
\begin{eqnarray}S_A\lineup = \int_0^1 dt\, \sum_{n=0}^\infty \omega_S(\PsiA,M_{n+1}(t\PsiA,...,t\PsiA))\nonumber\\
\lineup = \int_0^1 dt\, \omega_S\left(\PsiA, \pi_1\M\frac{1}{1-t\PsiA}\right).\label{eq:SA2}\end{eqnarray}
Next we make use the fact that the cohomomorphism $\G$ is {\it cyclic}. A cyclic cohomomorphism is one which can be obtained as a path-ordered exponential of coderivations representing cyclic multi-string products. The relevant property for us is that a cohomomorphism $\H$ which is cyclic with respect to a symplectic form $\omega$ satisfies
\begin{equation}\omega\left(\pi_1 \C \frac{1}{1-A},\pi_1\D\frac{1}{1-A}\right)=\omega\left(\pi_1 \H\C \frac{1}{1-A},\pi_1\H\D\frac{1}{1-A}\right).\label{eq:Okid}\end{equation}
where $A$ is a string field and $\C$ and $\D$ are arbitrary coderivations. For a proof see appendix A of \cite{OkWB}. In preparation for applying this identity, note that we can write \eq{SA2} in the form
\begin{equation}S_A = \int_0^1 dt\, \frac{1}{t}\omega_S\left(\pi_1{\bf 1}\frac{1}{1-t\PsiA}, \pi_1\M\frac{1}{1-t\PsiA}\right),\label{eq:SgA2}\end{equation}
where ${\bf 1}$ is the coderivation derived from the identity operator $\mathbb{I}$, regarded as a 1-string product. Explicitly
\begin{equation}{\bf 1} = \sum_{n=1}^\infty n \pi_{n}.\end{equation}
Now since by assumption $\G$ is in the small Hilbert space, we can apply \eq{Okid} together with $\G\M = \Q\G$ to find
\begin{equation}S_A = \int_0^1 dt\, \frac{1}{t}\omega_S\left(\pi_1\G {\bf 1}\frac{1}{1-t\PsiA}, \pi_1\Q\G \frac{1}{1-t\PsiA}\right).\end{equation}
Now note that 
\begin{eqnarray}{\bf 1}\frac{1}{1-t\PsiA} \lineup = \sum_{n=1}^\infty n\underbrace{(t\PsiA)\otimes...\otimes(t\PsiA)}_{n\ \mathrm{times}}, \nonumber\\
\lineup =\sum_{n=1}^\infty nt^n \underbrace{\PsiA\otimes...\otimes\PsiA}_{n\ \mathrm{times}},\nonumber\\
\lineup = t\frac{d}{dt}\sum_{n=1}^\infty t^n \underbrace{\PsiA\otimes...\otimes\PsiA}_{n\ \mathrm{times}},\nonumber\\
\lineup = t\frac{d}{dt}\frac{1}{1-t\PsiA},
\end{eqnarray}
so the action further simplifies to 
\begin{eqnarray}S_A \lineup = \int_0^1 dt\, \omega_S\left(\frac{d}{dt}\pi_1\G \frac{1}{1-t\PsiA}, \pi_1\Q\G \frac{1}{1-t\PsiA}\right)\nonumber\\
\lineup = \int_0^1 dt\, \omega_S\left(\frac{d}{dt}G[t\PsiA], Q G[t\PsiA]\right) .
\end{eqnarray}
Now note 
\begin{eqnarray}S_A 
\lineup = \int_0^1 dt\, \frac{d}{dt}\Big(\omega_S(G[t\PsiA], Q G[t\PsiA])\Big) - \int_0^1 dt\, \omega_S\left(G[t\PsiA], Q \frac{d}{dt}G[t\PsiA]\right),\nonumber\\
\lineup = \omega_S\left(G[\PsiA], Q G[\PsiA]\right) + \int_0^1 dt\, \omega_S\left(Q G[t\PsiA],\frac{d}{dt}G[t\PsiA]\right),\nonumber\\
\lineup = \omega_S\left(G[\PsiA], Q G[\PsiA]\right) - \int_0^1 dt\, \omega_S\left(\frac{d}{dt} G[t\PsiA],QG[t\PsiA]\right),\nonumber\\
\lineup = \omega_S\left(G[\PsiA], Q G[\PsiA]\right) -S_A.\end{eqnarray}
Therefore we find 
\begin{equation}S_A = \frac{1}{2}\omega_S\left(G[\PsiA], Q G[\PsiA]\right).\end{equation}
When $\xi$ is in the small Hilbert space, the $A_\infty$ action is related to a free action by field redefinition.

\subsection{Mapping the Reduced Berkovits Action into a Free Action}
\label{subsec:Bfree}

Next, we want to show that the reduced Berkovits action can be mapped to a free action when $\xi$ is in the small Hilbert space. Let us start with
\begin{equation}S_B=-\frac{1}{g_o^2}\int_0^1 dt\,\left\langle \xi\PsiB,Q\Big((\eta e^{t\xi\PsiB}) e^{-t\xi\PsiB}\Big)\right\rangle_L \end{equation}
The action in this form is not well-defined when $\xi$ is in the small Hilbert space. We first have to eliminate $\eta$ and $g_o$ and replace the large Hilbert space BPZ inner product with the small Hilbert space BPZ inner product. To simplify the appearance of subsequent formulas, denote
\begin{equation}g^t  \equiv e^{t\xi\PsiB}.\end{equation}
We have  
\begin{equation}\eta g^t = g_o \int_0^t ds\, g^s\, \PsiB\, g^{1-s}.\end{equation}
Thus we can rewrite the action 
\begin{equation}S_B=-\frac{1}{g_o}\int_0^1 dt\,\int_0^t ds\, \left\langle \xi\PsiB,Q\Big(g^s\,\PsiB\,g^{-s}\Big)\right\rangle_L .\end{equation}
Switching the order of the integration over $s$ and $t$, 
\begin{equation}\int_0^1dt\, \int_0^t ds\, = \int_0^1 ds\,\int_t^1dt ,\end{equation}
and integrating out $t$ gives 
\begin{equation}S_B=-\frac{1}{g_o}\int_0^1 ds\, (1-s)\left\langle \xi\PsiB,Q\Big(g^s\,\PsiB\,g^{-s}\Big)\right\rangle_L .\end{equation}
Now project the second entry of the BPZ inner product into the small Hilbert space by inserting a trivial factor $[\eta,\xi]=g_o$:
\begin{equation}S_B=-\frac{1}{g_o^2}\int_0^1 ds\, (1-s)\left\langle \xi\PsiB,\eta\xi Q\Big(g^s\,\PsiB\,g^{-s}\Big)\right\rangle_L .\end{equation}
We can use \eq{BPZls} to replace the large Hilbert space BPZ inner product with the small Hilbert space BPZ inner product:
\begin{equation}S_B=\frac{1}{g_o}\int_0^1 ds\, (1-s)\left\langle \PsiB,\eta\xi Q\Big(g^s\,\PsiB\,g^{-s}\Big)\right\rangle_S .\end{equation}
Further acting $\eta$ on the second entry of the BPZ inner product gives 
\begin{equation}
S_B=\int_0^1 ds\, (1-s)\left\langle \PsiB, Q\Big(g^s\,\PsiB\,g^{-s}\Big)+\xi Q\left(\int_0^s dt\, [g^t\,\PsiB\,g^{-t},g^s\,\PsiB\,g^{-s}]\right)\right\rangle_S 
.\end{equation}
Now it is consistent to imagine that $\xi$ is in the small Hilbert space, and attempt to write this as a free action.

Perform the following manipulations:
\begin{eqnarray}
S_B\lineup =\int_0^1 ds\, (1-s)\left\langle \PsiB, Q\Big(g^s\,\PsiB\,g^{-s}\Big)\right\rangle_S\nonumber\\
\lineup\ \ \ +\int_0^1 ds\, \int_0^s dt\,(1-s)\Big\langle\xi \PsiB,  [Q(g^t\,\PsiB\,g^{-t}),g^s\,\PsiB\,g^{-s}] - [g^t\,\PsiB\,g^{-t},Q(g^s\,\PsiB\,g^{-s})]\Big\rangle_S, \nonumber\\
\lineup = \int_0^1 ds\, (1-s)\left\langle \PsiB, Q\Big(g^s\,\PsiB\,g^{-s}\Big)\right\rangle_S\nonumber\\
\lineup\ \ \ +\int_0^1 dt\, \int_t^1 ds\,(1-s)\left\langle [\xi \PsiB, g^s\,\PsiB\,g^{-s}],  Q\Big(g^t\,\PsiB\,g^{-t}\Big)\right\rangle_S\nonumber\\
\lineup\ \ \ +\int_0^1 ds\, \int_0^s dt\,(1-s)\left\langle [\xi \PsiB, g^t\,\PsiB\,g^{-t}],  Q\Big(g^s\,\PsiB\,g^{-s}\Big)\right\rangle_S.
\end{eqnarray}
In the first equation we acted $Q$ on the commutator and moved $\xi$ to the first entry of the BPZ inner product. In the second equation we moved the commutator onto the first entry of the BPZ inner product, and in one term switched the order of integration of $s$ and $t$. Next note that 
\begin{equation}\frac{d}{ds}g^s = g^s (\xi\PsiB) = (\xi\PsiB) g^s,\end{equation}
and in particular
\begin{equation}\frac{d}{ds} \Big(g^s\, \PsiB\, g^{-s}\Big) = [\xi\PsiB, g^s\, \PsiB\, g^{-s}].\end{equation}
Then we have
\begin{eqnarray}
S_B \lineup = \int_0^1 ds\, (1-s)\left\langle \PsiB, Q\Big(g^s\,\PsiB\,g^{-s}\Big)\right\rangle_S\nonumber\\
\lineup\ \ \ +\int_0^1 dt\, \int_t^1 ds\,(1-s)\frac{d}{ds}\left\langle g^s\,\PsiB\,g^{-s},  Q\Big(g^t\,\PsiB\,g^{-t}\Big)\right\rangle_S\nonumber\\
\lineup\ \ \ +\int_0^1 ds\, \int_0^s dt\,(1-s)\frac{d}{dt}\left\langle  g^t\,\PsiB\,g^{-t},  Q\Big(g^s\,\PsiB\,g^{-s}\Big)\right\rangle_S,\nonumber\\
\lineup =  \int_0^1 ds\, (1-s)\left\langle \PsiB, Q\Big(g^s\,\PsiB\,g^{-s}\Big)\right\rangle_S\nonumber\\
\lineup\ \ \ +\int_0^1 dt\, \int_t^1 ds\,\frac{d}{ds}\left((1-s)\left\langle g^s\,\PsiB\,g^{-s},  Q\Big(g^t\,\PsiB\,g^{-t}\Big)\right\rangle_S\right)
\nonumber\\
\lineup\ \ \ +\int_0^1 dt\, \int_t^1 ds\,\left\langle g^s\,\PsiB\,g^{-s},  Q\Big(g^t\,\PsiB\,g^{-t}\Big)\right\rangle_S\nonumber\\
\lineup\ \ \ +\int_0^1 ds\,(1-s)\left\langle  g^s\,\PsiB\,g^{-s},  Q\Big(g^s\,\PsiB\,g^{-s}\Big)\right\rangle_S\nonumber\\
\lineup\ \ \ -\int_0^1 ds\,(1-s)\left\langle\PsiB,  Q\Big(g^s\,\PsiB\,g^{-s}\Big)\right\rangle_S.
\end{eqnarray}
In the second step we pulled $d/ds$ through the $(1-s)$ factor and integrated the total $d/dt$ derivative. Canceling terms and integrating the total $d/ds$ derivative,
\begin{eqnarray}
S_B \lineup = \int_0^1 dt\, \int_t^1 ds\,\frac{d}{ds}\left((1-s)\left\langle g^s\,\PsiB\,g^{-s},  Q\Big(g^t\,\PsiB\,g^{-t}\Big)\right\rangle_S\right)
\nonumber\\
\lineup\ \ \ +\int_0^1 dt\, \int_t^1 ds\,\left\langle g^s\,\PsiB\,g^{-s},  Q\Big(g^t\,\PsiB\,g^{-t}\Big)\right\rangle_S\nonumber\\
\lineup\ \ \ +\int_0^1 dt\,(1-t)\left\langle  g^t\,\PsiB\,g^{-t},  Q\Big(g^t\,\PsiB\,g^{-t}\Big)\right\rangle_S,\nonumber\\
\lineup = \int_0^1 dt\, \int_t^1 ds\,\left\langle g^s\,\PsiB\,g^{-s},  Q\Big(g^t\,\PsiB\,g^{-t}\Big)\right\rangle_S.
\end{eqnarray}
Symmeterizing the order of integration and relabeling $s$ and $t$,
\begin{eqnarray}
S_B \lineup = \frac{1}{2}\left (\int_0^1 dt\, \int_t^1 ds\,+\int_0^1 ds\, \int_0^s dt\,\right)\left\langle g^s\,\PsiB\,g^{-s},  Q\Big(g^t\,\PsiB\,g^{-t}\Big)\right\rangle_S,\nonumber\\
\lineup = \frac{1}{2} \int_0^1 dt\, \int_t^1 ds\,\left\langle g^s\,\PsiB\,g^{-s},  Q\Big(g^t\,\PsiB\,g^{-t}\Big)\right\rangle_S
+\frac{1}{2}\int_0^1 dt\, \int_0^t ds\,\left\langle g^t\,\PsiB\,g^{-t},  Q\Big(g^s\,\PsiB\,g^{-s}\Big)\right\rangle_S,\nonumber\\
\lineup =\frac{1}{2}\left (\int_0^1 dt\, \int_t^1 ds\,+\int_0^1 dt\, \int_0^t ds\,\right)\left\langle g^s\,\PsiB\,g^{-s},  Q\Big(g^t\,\PsiB\,g^{-t}\Big)\right\rangle_S,\nonumber\\
\lineup = \frac{1}{2} \left\langle \int_0^1 ds \,g^s\,\PsiB\,g^{-s},  \int_0^1 dt\, Q\Big(g^t\,\PsiB\,g^{-t}\Big)\right\rangle_S.
\end{eqnarray}
Therefore
\begin{equation}S_B = \frac{1}{2}\big\langle F[\PsiB],QF[\PsiB]\big\rangle_S.\end{equation}
When $\xi$ is in the small Hilbert space, the reduced Berkovits action is related to a free action by field redefinition.

\section{Concluding Remarks}
\label{sec:conclusion}

In this paper we have shown that the $A_\infty$ superstring field theory is related to a free theory through the improper field redefinition given in \eq{Gexp}
\begin{equation}\Psi_0 = G[\PsiA].\end{equation}
Likewise, the reduced Berkovits theory is related to a free theory through the improper field redefinition given in \eq{Fexp}
\begin{equation}\Psi_0= F[\PsiB].\end{equation}
The field redefinition between the $A_\infty$ and reduced Berkovits theories is then given by equating
\begin{equation}G[\PsiA] = F[\PsiB].\end{equation}
It is worth mentioning that this result does not depend on any deep structural connection between the $A_\infty$ and reduced Berkovits theories. Our analysis shows that it is generically true that any pair of superstring field theories in the small Hilbert space can be related by equating respective improper field redefinitions to a free theory. For the reduced Berkovits and $A_\infty$ theories, the improper field redefinitions $G[\PsiA]$ and $F[\PsiB]$ have no obvious relation beyond being constructed out of $\xi$ and the open string star product.

The fact that we have not uncovered such a relation, however, does not imply that one does not exist. One particularly remarkable coincidence is that the products entering the field redefinition up to cubic order have no cyclic component:
\begin{eqnarray}
E_0\lineup =0,\nonumber\\
\frac{1}{2}\Big(1+\omega\Big)\circ E_1 \lineup = 0,\nonumber\\
\frac{1}{3}\Big(1+\omega+\omega^2\Big)\circ E_2 \lineup = 0,\nonumber\\
\frac{1}{4}\Big(1+\omega+\omega^2+\omega^3\Big)\circ E_3 \lineup = 0.
\end{eqnarray}
The fact that this holds for $E_0$ and $E_1$ is trivial, and that it holds for $E_2$ is almost trivial---it follows for any field redefinition in the small Hilbert space built from $\xi$ and $m_2$. That this holds for $E_3$ is highly nontrivial. Assuming vertices built from $Q,m_2$ and $\xi$, there is a two dimensional space of theories realizing cyclic $A_\infty$ at quartic order. The $A_\infty$ superstring field theory is the unique one that can be obtained from the reduced Berkovits theory by a field redefinition with no cyclic component.  It would be interesting to see if this correspondence holds to higher orders.

\bigskip 

\noindent {\bf Acknowledgments} 

\bigskip

\noindent The author would like to thank Y. Okawa and T. Takezaki for collaboration, and S. Konopka for discussions of  mathematical apparatus. This work was supported in parts by the DFG Transregional Collaborative Research Centre TRR 33, and the DFG cluster of excellence Origin and Structure of the Universe.

\begin{appendix}

\section{The Coproduct}
\label{app:coproduct}

The tensor algebra has a naturally defined {\it coproduct}, which is a linear operation which takes an object in the tensor algebra and produces an object in 
a {\it pair} of tensor algebras:
\begin{equation}\triangle: T\mathcal{H}\, \to\, T\mathcal{H}\otimes' T\mathcal{H}.\end{equation}
Note that the tensor product $\otimes'$ (with a prime) is different from the tensor product $\otimes$ which defines $T\mathcal{H}$. The coproduct is the reverse, or the ``dual" of a product. Acting on factorized states in the tensor algebra, the coproduct gives
\begin{eqnarray}
\triangle 1_{T{\mathcal{H}}}\lineup = 1_{T{\mathcal{H}}}\otimes' 1_{T{\mathcal{H}}},\nonumber\\
\triangle A \lineup = 1_{T{\mathcal{H}}}\otimes' A + A\otimes' 1_{T{\mathcal{H}}},\nonumber\\
\triangle (A\otimes B) \lineup = 1_{T{\mathcal{H}}}\otimes' (A\otimes B) + A\otimes' B+(A\otimes B)\otimes'1_{T{\mathcal{H}}},\nonumber\\
\triangle (A\otimes B\otimes C) \lineup = 1_{T{\mathcal{H}}}\otimes'(A\otimes B\otimes C) + A\otimes'(B\otimes C) +(A\otimes B)\otimes' C + (A\otimes B\otimes C)\otimes'1_{T{\mathcal{H}}}.\ \ \ \ \ \ 
\end{eqnarray}
More generally,
\begin{equation}
\triangle(A_1\otimes...\otimes A_n) = 
\sum_{k=0}^{n} (A_0\otimes A_1\otimes ...\otimes A_k)\otimes'(A_{k+1}\otimes...\otimes A_n\otimes A_{n+1}),
\end{equation}
where the $A_0=A_{n+1}=1_{T\mathcal{H}}$. Note that $1_{T{\mathcal{H}}}$ is not regarded as the identity with respect to the primed tensor product $\otimes'$. The upshot is that the coproduct sums over all distinct ways of replacing the tensor product $\otimes$ in an element of the tensor algebra with a primed tensor product $\otimes'$. One can check that the coproduct is {\it coassociative}:
\begin{equation}(\triangle\otimes'\mathbb{I}_{T\mathcal{H}})\triangle = (\mathbb{I}_{T\mathcal{H}}\otimes'\triangle)\triangle.\end{equation}
This allows us to unambiguously define repeated coproducts. Besides the coproduct, the tensor algebra has a natural notion of multiplication defined by the tensor product $\otimes$ itself. It is convenient to describe this operation using a linear operator $\inverttriangle$, which maps two copies of the tensor algebra into one copy:
\begin{equation}\inverttriangle:T\mathcal{H}\otimes'T\mathcal{H}\to T\mathcal{H}.\end{equation}
The operation $\inverttriangle$ acts by replacing the primed tensor product $\otimes'$ in the input with the tensor product $\otimes$. The multiplication $\inverttriangle$ is associative:
\begin{equation}\inverttriangle\!(\mathbb{I}_{T\mathcal{H}}\otimes'\inverttriangle ) = \inverttriangle\!(\inverttriangle\!\otimes'\mathbb{I}_{T\mathcal{H}}).\end{equation}
This allows us to unambiguously define repeated products.

By definition, a {\it coderivation} is a linear operator on the tensor algebra which satisfies 
\begin{equation}\triangle \D = (\D\otimes'\mathbb{I}_{T\mathcal{H}} +\mathbb{I}_{T\mathcal{H}}\otimes'\D).\label{eq:cD}\triangle\end{equation}
This property can be interpreted as ``dual" to the Leibniz product rule. By definition, a {\it cohomomorphism} is a linear operator of even degree on the tensor algebra which satisfies 
\begin{equation}\triangle\H = (\H\otimes'\H).\label{eq:cH}\triangle \end{equation}
This can be interpreted as ``dual" to the defining property of an algebra homomorphism. A {\it group-like element} of the tensor algebra $V\in T\mathcal{H}$ satisfies 
\begin{equation}\triangle V = V\otimes' V.\label{eq:cgl}\end{equation}
We will see how these definitions relate to the explicit formulas \eq{D}, \eq{H}, and \eq{gl} in a moment. For now we note the following useful properties, which follow immediately from the definitions: 

\bigskip

\begin{Fact}The product of two cohomomorphisms is a cohomomorphism.\end{Fact}

\begin{Fact}The commutator of two coderivations (graded with respect to degree) is a coderivation.\end{Fact}

\begin{Fact} If a cohomomorphism has an inverse operator on the tensor algebra, the inverse operator is also a cohomomorphism.\end{Fact}

\begin{Fact}If $\D$ is a coderivation and $\H$ an invertible cohomomorphism, $\H^{-1}\D\H$ is a coderivation.\end{Fact}

\begin{Fact} Given a 1-parameter family of degree even coderivations, their path-ordered exponential is a cohomomorphism.\end{Fact} 

\bigskip

\noindent To see property 1, note that if $\H_1$ and $\H_2$ are two cohomomorphisms then
\begin{equation}\triangle \H_1\H_2 = (\H_1\otimes'\H_1)\triangle\H_2 =(\H_1\otimes'\H_1)(\H_2\otimes'\H_2)\triangle  = 
((\H_1\H_2)\otimes'(\H_1\H_2)),\triangle\end{equation}
which means that $\H_1\H_2$ is a cohomomorphism. To see property 5, consider for example the operator $\G(t_1,t_2)$ introduced in \eq{Gt1t2}, which can be defined as a limit
\begin{equation}\G(t_1,t_2)=\lim_{N\to\infty}\left(\mathbb{I}_{T\mathcal{H}}+\frac{\mmu(t_1)}{N}\right)
\left(\mathbb{I}_{T\mathcal{H}}+\frac{\mmu\left(t_1\!+\!\fraction{t_2-t_1}{N}\right)}{N}\right)
\left(\mathbb{I}_{T\mathcal{H}}+\frac{\mmu\left(t_1\!+\!2\fraction{t_2-t_1}{N}\right)}{N}\right)...
\left(\mathbb{I}_{T\mathcal{H}}+\frac{\mmu(t_2)}{N}\right).
\end{equation} 
To order $1/N$, each factor $(\mathbb{I}_{T\mathcal{H}}+\mmu(t)/N)$ is a cohomomorphism. Therefore, by property 1, the product of all of these factors is also a cohomomorphism.

We would like to prove the explicit formulas \eq{D}, \eq{H} and \eq{gl} starting from the definition of coderivations, cohomomorphisms, and group-like elements in terms of the coproduct. For this we need two identities:
\begin{equation}
\pi_{m+n}= \inverttriangle\! (\pi_m\otimes' \pi_n)\triangle,\label{eq:cid1}\end{equation}
and
\begin{equation}
\inverttriangle\! ((b_{k,m}\pi_m)\otimes' (c_{\ell,n}\pi_{n}))  = (b_{k,m}\otimes c_{\ell,n})\inverttriangle \!(\pi_m\otimes'\pi_n).\label{eq:cid2}
\end{equation}
where $b_{k,m}$ is a linear map from $\mathcal{H}^{\otimes m}$ into $\mathcal{H}^{\otimes k}$ and $c_{\ell,n}$ is a linear map from $\mathcal{H}^{\otimes n}$ into $\mathcal{H}^{\otimes \ell}$. The first identity works as follows. The coproduct $\triangle$ splits the input in all possible ways into two copies of the tensor algebra, but, because the coproduct is followed by $\pi_m\otimes'\pi_n$, only the unique term which is an $m$-string state in the first copy an an $n$-string state in the second copy will make a contribution. The product $\inverttriangle$ then joins these states together, and the net effect is a projection onto the $m+n$-string component of the tensor algebra. The second identity follows since, acting on a state in $\mathcal{H}^{\otimes m}\otimes'\mathcal{H}^{\otimes n}$, it does not matter whether the operators $b_{k,m}$ and $c_{\ell,n}$ act before or after $\inverttriangle$ replaces the primed tensor product $\otimes'$ with the tensor product $\otimes$. Plugging
\eq{cid1} back into itself recursively gives the identity
\begin{equation}\pi_{k_1+k_2+...+k_\ell} = \inverttriangle^{\ell} \!(\pi_{k_1}\otimes'\pi_{k_2}\otimes'...\otimes'\pi_{k_\ell})\triangle^\ell,\end{equation}
where $\inverttriangle^\ell$ and $\triangle^\ell$ denote $\ell$-times repeated products and coproducts, respectively. This can be defined, for example, recursively by the formulas
\begin{equation}
\inverttriangle^{\ell+1} \equiv\inverttriangle^\ell (\inverttriangle\!\otimes' \underbrace{\mathbb{I}_{T\mathcal{H}}\otimes' ... \otimes' \mathbb{I}_{T\mathcal{H}}}_{\ell\ \mathrm{times}}),\ \ \ \ \ \ \triangle^{\ell+1} \equiv (\triangle\otimes' \underbrace{\mathbb{I}_{T\mathcal{H}}\otimes' ... \otimes' \mathbb{I}_{T\mathcal{H}}}_{\ell\ \mathrm{times}})\triangle^\ell.
\end{equation}
The ordering of $\triangle$ and $\inverttriangle$ with respect to the identity maps on the right hand side is not essential since the product is associative and the coproduct is coassociative. From the definitions \eq{cD} and \eq{cH} it is easy to see that coderivations and cohomomorphisms satisfy
\begin{equation}\triangle^{\ell+1} \D = \Big(\D\otimes'\underbrace{\mathbb{I}_{T\mathcal{H}}\otimes'...\otimes'\mathbb{I}_{T\mathcal{H}}}_{\ell\ \mathrm{times}}\, +\, \mathbb{I}_{T\mathcal{H}}\otimes'\D\otimes'\underbrace{\mathbb{I}_{T\mathcal{H}}\otimes'...\otimes'\mathbb{I}_{T\mathcal{H}}}_{\ell-1\ \mathrm{times}}\,+\,...\,+\,\underbrace{\mathbb{I}_{T\mathcal{H}}\otimes'...\otimes'\mathbb{I}_{T\mathcal{H}}}_{\ell\ \mathrm{times}}\otimes' \D\Big)\triangle^n,\ \ \ \ \ \end{equation}
and 
\begin{equation}
\triangle^\ell \H = \Big(\underbrace{\H\otimes'...\otimes'\H}_{\ell\ \mathrm{times}}\Big)\triangle^n.
\end{equation}
Note the resemblance to \eq{D} and \eq{H}. Now we have the ingredients needed to prove \eq{D}, \eq{H} and \eq{gl}.

Let's start with the group-like element $V$. Projecting onto the 1-string component we get a string field we can call $\Psi$:
\begin{equation}\Psi \equiv \pi_1 V.\end{equation}
Projecting on the 2-string component we can use \eq{cid1} to find
\begin{eqnarray}\pi_2 V\lineup  = \inverttriangle\! (\pi_1\otimes'\pi_1) \triangle V,\nonumber\\
\lineup = \inverttriangle\! (\pi_1\otimes'\pi_1) (V\otimes' V),\nonumber\\
\lineup = \inverttriangle\! ((\pi_1 V)\otimes'(\pi_1V)),\nonumber\\
\lineup = \inverttriangle\! (\Psi \otimes'\Psi),\nonumber\\
\lineup = \Psi\otimes\Psi.
\end{eqnarray}
Similarly we find $\pi_3 V = \Psi\otimes\Psi\otimes\Psi$ and so on. To get the zero string component, consider the equation
\begin{eqnarray}\Psi \lineup = \pi_1 V,\nonumber\\
\lineup  = \inverttriangle\! (\pi_0\otimes'\pi_1) \triangle V,\nonumber\\
\lineup = \inverttriangle\! (\pi_0 V \otimes'\Psi),\nonumber\\
\lineup = (\pi_0 V)\otimes \Psi.\end{eqnarray}
This is only consistent if $\pi_0 V = 1_{T\mathcal{H}}$. Therefore
\begin{equation}V = \sum_{n=0}^{\infty} \pi_n V = 1_{T\mathcal{H}}+ \sum_{n=0}^{\infty}\Psi^{\otimes n+1} = \frac{1}{1-\Psi}.\end{equation}
in agreement with \eq{gl}. 

Now we can apply a similar procedure to find the general form of a coderivation $\D$. The action of $\pi_1\D$ on an $m$-string state naturally defines an 
$m$-string product
\begin{equation}\pi_1\D (A_1\otimes...\otimes A_m) \equiv D_m(A_1,...,A_m).\end{equation}
Therefore we can write
\begin{equation}\pi_1\D = \sum_{m=0}^\infty D_m\pi_m.\label{eq:Dprod}\end{equation}
Compute the projection on the 2-string component,
\begin{eqnarray}
\pi_2 \D \lineup = \inverttriangle\! (\pi_1\otimes'\pi_1) \triangle\D,\nonumber\\
\lineup  = \inverttriangle\! (\pi_1\otimes'\pi_1)(\D\otimes ' \mathbb{I}_{T\mathcal{H}}+\mathbb{I}_{T\mathcal{H}}\otimes' \D)\triangle,\nonumber\\
\lineup = \inverttriangle\!((\pi_1\D)\otimes ' \pi_1+\pi_1\otimes' (\pi_1\D))\triangle\nonumber,\\
\lineup = \sum_{m=0}^\infty \inverttriangle\!((D_m\pi_m)\otimes ' \pi_1+\pi_1\otimes' (D_m\pi_m))\triangle,\nonumber\\
\lineup = \sum_{m=0}^\infty \Big[(D_m\otimes\mathbb{I})\inverttriangle\!(\pi_m\otimes ' \pi_1)\triangle +(\mathbb{I}\otimes D_m)\inverttriangle\!(\pi_1\otimes' \pi_m)\triangle\Big],\nonumber\\
\lineup =\sum_{m=0}^\infty (D_m\otimes\mathbb{I}+\mathbb{I}\otimes D_m)\pi_{m+1}.
\end{eqnarray}
In the first step we used the identity \eq{cid1}, in the second step we used the coderivation property of $\D$, in the fourth step we plugged in \eq{Dprod}, in the fifth step we used \eq{cid2}, and the sixth step we again used \eq{cid1}. Similarly computing 
\begin{equation}\pi_3 \D = \inverttriangle^2 \!\!(\pi_1\otimes'\pi_1\otimes'\pi_1)\triangle^2 \D, \end{equation}
we find 
\begin{equation}\pi_3\D =\sum_{m=0}^\infty (D_m\otimes\mathbb{I}\otimes\mathbb{I}+\mathbb{I}\otimes D_m\otimes\mathbb{I}+\mathbb{I}\otimes\mathbb{I}\otimes D_m)\pi_{m+2}.\end{equation}
and so on. This leaves us to compute the projection onto the zero-string component of $\D$. In general, such a projection can take the form
\begin{equation}\pi_0\D = \sum_{m=0}^\infty d_m \pi_m,\end{equation}
where $d_m$ is a map from $\mathcal{H}^{\otimes m}$ into $\mathcal{H}^{\otimes 0}$. To see what the $d_m$s must be, compute
\begin{eqnarray}\pi_1\D \lineup = \inverttriangle\! (\pi_0\otimes' \pi_1)\triangle \D, \nonumber\\
\lineup = \inverttriangle\! ((\pi_0\D)\otimes' \pi_1 + \pi_0\otimes' (\pi_1\D))\triangle,\nonumber\\
\lineup = \sum_{m=0}^\infty\inverttriangle\! ((d_m \pi_m)\otimes' \pi_1 + \pi_0\otimes' (D_m\pi_m))\triangle,\nonumber\\
\lineup = \sum_{m=0}^\infty\Big[(d_m\otimes\mathbb{I})\inverttriangle\! (\pi_m\otimes' \pi_1)\triangle + D_m\inverttriangle\!(\pi_0\otimes'\pi_m)\triangle\Big],\nonumber\\
\lineup = \sum_{m=0}^\infty\Big[(d_m\otimes\mathbb{I})\pi_{m+1} + D_m\pi_m\Big],\nonumber\\
\lineup = \sum_{m=0}^\infty(d_m\otimes\mathbb{I})\pi_{m+1}+\pi_1\D.\end{eqnarray}
from which we learn that $d_m=0$ and $\pi_0\D = 0$. Therefore a coderivation must take the general form
\begin{eqnarray}
\D \lineup = \sum_{n=0}^\infty \pi_{n} \D,\nonumber\\
\lineup = \sum_{m=0}^{\infty} \sum_{n=0}^\infty\sum_{k=0}^n (\mathbb{I}^{\otimes k} \otimes D_m \otimes \mathbb{I}^{\otimes n-k}) \pi_{m+n},\nonumber\\
\lineup = \sum_{\ell,m,n=0}^{\infty}(\mathbb{I}^{\otimes \ell} \otimes D_m \otimes \mathbb{I}^{\otimes n}) \pi_{\ell+m+n}.
\end{eqnarray}
which reproduces the expression \eq{D}.

Finally, let's consider cohomomorphisms. We can take
\begin{equation}\pi_1\H = \sum_{k=0}^\infty H_k\pi_k.\label{eq:Hprod}\end{equation}
where $H_k$ are a sequence of degree even multi-string products. Compute the projection onto the 2-string component
\begin{eqnarray}\pi_2 \H \lineup = \inverttriangle\! (\pi_1\otimes'\pi_1)\triangle \H,\nonumber\\
\lineup = \inverttriangle\! (\pi_1\H)\otimes'(\pi_1\H)\triangle,\nonumber\\
\lineup =\sum_{k_1,k_2=0}^{\infty} \inverttriangle\! (H_{k_1}\pi_{k_1})\otimes'(H_{k_2}\pi_{k_2})\triangle,\nonumber\\
\lineup =\sum_{k_1,k_2=0}^{\infty} (H_{k_1}\otimes H_{k_2})\inverttriangle\! (\pi_{k_1}\otimes'\pi_{k_2})\triangle,\nonumber\\
\lineup =\sum_{k_1,k_2=0}^{\infty} (H_{k_1}\otimes H_{k_2})\pi_{k_1+k_2}.
\end{eqnarray}
Similarly for the 3-string component we compute 
\begin{equation}\pi_3 \H = \inverttriangle^2 \!\!(\pi_1\otimes'\pi_1\otimes'\pi_1)\triangle^2 \H \end{equation}
to find 
\begin{equation}\pi_3\H = \sum_{k_1,k_2,k_3=0}^{\infty} (H_{k_1}\otimes H_{k_2}\otimes H_{k_3})\pi_{k_1+k_2+k_3},\end{equation}
and so on. Next we must compute the projection onto the zero-string component. In general it can take the form
\begin{equation}\pi_0\H = \sum_{k=0}^{\infty} h_k\pi_k,\end{equation}
where $h_k$ are maps from $\mathcal{H}^{\otimes k}$ to $\mathcal{H}^{\otimes 0}$. To see what the $h_k$s must be, compute
\begin{eqnarray}
\pi_0\H \lineup = \inverttriangle\! (\pi_0\otimes'\pi_0)\triangle \H,\nonumber\\
\lineup =  \inverttriangle\! ((\pi_0\H)\otimes'(\pi_0\H))\triangle,\nonumber\\
\lineup = \sum_{k_1,k_2=0}^\infty (h_{k_1}\otimes h_{k_2}) \inverttriangle\! (\pi_{k_1}\otimes'\pi_{k_2})\triangle,\nonumber\\
\lineup = \sum_{k_1,k_2=0}^\infty (h_{k_1}\otimes h_{k_2}) \pi_{k_1+k_2}.
\end{eqnarray}
Comparing to \eq{Hprod} implies
\begin{equation}h_k = \sum_{i=0}^k h_i\otimes h_{k-i},\label{eq:HhH}\end{equation}
or
\begin{eqnarray}
h_0 \lineup = h_0\otimes h_0,\\
h_1 \lineup = h_0\otimes h_1 + h_1\otimes h_0,\\
h_2 \lineup = h_0\otimes h_2 + h_1\otimes h_1 + h_2\otimes h_0.\\
\lineup \vdots\ .
\end{eqnarray}
From this it is clear that either $h_0=0$, or $h_0 = \mathbb{I}^{\otimes 0}$ (the identity operator on $\mathcal{H}^{\otimes 0}$). In both cases $h_k=0$ for $k\geq 1$. To further distinguish the two possibilities for $h_0$, compute
\begin{eqnarray}
\pi_1\H \lineup = \inverttriangle\! (\pi_0\otimes'\pi_1)\triangle \H,\nonumber\\
\lineup =  \inverttriangle\! ((\pi_0\H)\otimes'(\pi_1\H))\triangle,\nonumber\\
\lineup = \sum_{k=0}^\infty (h_0\otimes H_{k}) \pi_{k}.
\end{eqnarray}
Comparing to \eq{Hprod}
\begin{equation}H_k = h_0\otimes H_{k}.\label{eq:HhH2}\end{equation}
We can only have $h_0 = 0$ if all multi-string products vanish, and in this case the entire cohomomorphism vanishes: $\H=0$. If the multi-string products do not vanish, we can only have $h_0=\mathbb{I}^{\otimes 0}$ and 
\begin{equation}\pi_0 \H = \pi_0.\end{equation}
Therefore a nonvanishing cohomomorphism takes the general form
\begin{eqnarray}
\H \lineup = \sum_{\ell=0}^{\infty}\pi_\ell \H,\nonumber\\
\lineup = \pi_0 + \sum_{\ell=1}^\infty\,\sum_{k_1=0}^\infty...\sum_{k_{\ell}=0}^\infty (H_{k_1}\otimes H_{k_2}\otimes ... \otimes H_{k_{\ell}})\pi_{k_1+k_2+...+k_{\ell}}. 
\end{eqnarray}
which reproduces the expression \eq{H}.

\section{Proof of \eq{Fn}}
\label{app:nF}

In this appendix we prove the identity \eq{Fn}:  
\begin{equation}\F\n\F^{-1} = \n - g_o \m_2.\end{equation}
It is convenient first to demonstrate a related formula satisfied by the products $F_{n+1}$: 
\begin{equation}[\eta,F_{n+2}] = g_o\sum_{k=0}^n m_2 (F_{k+1}\otimes F_{n-k+1}).\label{eq:etaF}\end{equation}
We prove this recursively. First assume that \eq{etaF} holds for some fixed $n$. Then we show that it holds for $n\to n+1$, as follows:
\begin{eqnarray}
[\eta,F_{n+3}] \lineup = -\frac{1}{n+3}[\eta, m_2(\xi\otimes F_{n+2}+F_{n+2}\otimes\xi)],\nonumber\\
\lineup =\frac{g_o}{n+3}\left[m_2(\mathbb{I}\otimes F_{n+2}+F_{n+2}\otimes\mathbb{I}) -\sum_{k=0}^n m_2(\xi\otimes m_2(F_{k+1}\otimes F_{n-k+1}))
\right. \nonumber\\
\lineup\ \ \ \ \ \ \ \ \ \ \ \ \ \left.+\sum_{k=0}^n m_2(m_2(F_{k+1}\otimes F_{n-k+1})\otimes \xi)\right],\nonumber\\
\lineup =\frac{g_o}{n+3}\left[m_2(\mathbb{I}\otimes F_{n+2}+F_{n+2}\otimes\mathbb{I}) -\sum_{k=0}^n m_2(m_2(\xi\otimes F_{k+1})\otimes F_{n-k+1}))
\right. \nonumber\\
\lineup\ \ \ \ \ \ \ \ \ \ \ \ \ \left.-\sum_{k=0}^n m_2(F_{k+1}\otimes m_2( F_{n-k+1}\otimes \xi))\right].\label{eq:nF1}
\end{eqnarray}
In the first step we use formula \eq{Frec}:
\begin{equation}F_{m+2} = -\frac{1}{m+2}m_2(\xi\otimes F_{m+1}+F_{m+1}\otimes \xi).\label{eq:Frecapp}\end{equation}
In the second step we computed the action of $\eta$, while assuming \eq{etaF} holds for our particular $n$. In the third step we used associativity of the star product to pull $m_2$ over $\xi$. Now in the second a third terms in \eq{nF1} add and subtract a piece where $\xi$ and $F$ are interchanged under $m_2$: 
\begin{eqnarray}
[\eta,F_{n+3}] \lineup =\frac{g_o}{n+3}\Big[m_2(\mathbb{I}\otimes F_{n+2}+F_{n+2}\otimes\mathbb{I}) \nonumber\\
\lineup\ \ \ -\sum_{k=0}^n m_2(m_2(\xi\otimes F_{k+1}+F_{k+1}\otimes \xi)\otimes F_{n-k+1})+\sum_{k=0}^n m_2(m_2(
F_{k+1}\otimes \xi)\otimes F_{n-k+1}))\nonumber\\
\lineup\ \ \  \left.-\sum_{k=0}^n m_2(F_{k+1}\otimes m_2( F_{n-k+1}\otimes \xi+\xi\otimes F_{n-k+1}))+\sum_{k=0}^n m_2(F_{k+1}\otimes m_2( \xi\otimes F_{n-k+1}))\right].\nonumber\\
\end{eqnarray}
In the second and fourth terms we can substitute \eq{Frecapp} for $F_{m+2}$, while the third and fifth terms cancel due to the associativity of $m_2$, leaving
\begin{eqnarray}
[\eta,F_{n+3}]\lineup =\frac{g_o}{n+3}\Big[m_2(\mathbb{I}\otimes F_{n+2}+F_{n+2}\otimes\mathbb{I}) \nonumber\\
\lineup\ \ \ \left.+\sum_{k=0}^n (k+2) m_2(F_{k+2}\otimes F_{n-k+1})+\sum_{k=0}^n (n-k+2)m_2(F_{k+1}\otimes F_{n-k+2})\right].
\end{eqnarray}
Relabeling indices, the second and third terms can be combined except for a mismatch at the upper and lower extremes of summation:
\begin{eqnarray}
[\eta,F_{n+3}] \lineup =\frac{g_o}{n+3}\Big[m_2(\mathbb{I}\otimes F_{n+2}+F_{n+2}\otimes\mathbb{I}) \nonumber\\
\lineup\ \ \ \left.+\sum_{k=1}^{n+1} (k+1) m_2(F_{k+1}\otimes F_{n-k+2})+\sum_{k=0}^n (n-k+2)m_2(F_{k+1}\otimes F_{n-k+2})\right],\nonumber\\
\lineup = \frac{1}{n+3}\left[m_2(\mathbb{I}\otimes F_{n+2}+F_{n+2}\otimes\mathbb{I}) +\sum_{k=1}^{n} (n+3) m_2(F_{k+1}\otimes F_{n-k+2})\right.
\nonumber\\ 
\lineup \ \ \ +(n+2)m_2(F_{n+2}\otimes F_1)+(n+2)m_2(F_1\otimes F_{n+2})\Big].
\end{eqnarray}
Recalling $F_1=\mathbb{I}$, the mismatched terms conveniently add to the first term, giving
\begin{equation}[\eta,F_{n+3}]  = g_o\sum_{k=0}^{n+1} m_2(F_{k+1}\otimes F_{n-k+2}),\end{equation}
which is \eq{etaF} with $n$ replaced by $n+1$. To complete the proof for all $n$, all that is left is to show that \eq{etaF} holds for $n=0$:
\begin{eqnarray}[\eta,F_2] \lineup = -\frac{1}{2}[\eta, m_2(\xi\otimes\mathbb{I}+\mathbb{I}\otimes\xi)],\nonumber\\
\lineup = g_o m_2, \nonumber\\
\lineup = g_o m_2(F_1\otimes F_1),
\end{eqnarray}
which it does.

We can return to the proof of \eq{Fn}. We can write \eq{Fn} in an equivalent form
\begin{equation}
 [\n,\F] - g_o \m_2\F = 0.
\end{equation}
To prove this formula, act the left hand side on the $(n+2)$-string component on the tensor algebra and project onto the 1-string component:
\begin{equation}
\pi_1\Big([\n,\F] - g_o \m_2\F\Big)\pi_{n+2} = \left([\eta,F_{n+2}] - g_o\sum_{k=0}^n m_2 (F_{k+1}\otimes F_{n-k+1})\right)\pi_{n+2} = 0.
\end{equation}
This vanishes as just demonstrated. We also have
\begin{eqnarray}
\pi_1\Big([\n,\F] - g_o \m_2\F\Big)\pi_0 \lineup =0,\\
\pi_1\Big([\n,\F] - g_o \m_2\F\Big)\pi_1 \lineup =0
\end{eqnarray}
because $F_0=0$ and $F_1=\mathbb{I}$ is in the small Hilbert space. Therefore
\begin{equation}\pi_1\Big([\n,\F] - g_o \m_2\F\Big) =0.\label{eq:pi1Fn}\end{equation}
Now let us compute the projection onto higher-string components. Using \eq{cid1} we can write 
\begin{eqnarray}
\pi_{n+1}\Big([\n,\F] - g_o \m_2\F\Big) \lineup = \inverttriangle \!(\pi_1\otimes ' \pi_n)\triangle \Big([\n,\F] - g_o \m_2\F\Big), \nonumber\\
\lineup = \inverttriangle \! \Big(\Big(\pi_1([\n,\F] - g_o \m_2\F)\Big)\otimes'(\pi_n\F)+(\pi_1\F)\otimes'\Big(\pi_n([\n,\F] - g_o \m_2\F)\Big)\Big)\triangle,\nonumber\\
\lineup = \inverttriangle \! \Big((\pi_1\F)\otimes'\Big(\pi_n ([\n,\F] - g_o \m_2\F)\Big)\Big)\triangle,
\end{eqnarray}
where in the last step we used \eq{pi1Fn}. This means that the projection onto the $(n+1)$-string component vanishes if the projection on the $n$-string component vanishes. We have already shown that the 1-string projection vanishes, so all $n$-string projections must vanish for $n\geq 1$. 
We have to check the zero-string projection separately: 
\begin{equation}\pi_0([\n,\F] - g_o \m_2\F)= -\pi_0\F\n=-\pi_0\n = 0.\end{equation}
where we use the fact that $\pi_0$ annihilates coderivations and returns $\pi_0$ on cohomomorphisms. This establishes \eq{Fn}.

\section{Cyclic Products of the Reduced Berkovits Theory}
\label{app:Bcyc}

In this appendix we show that the cyclic products $\tilde{B}_{n+1}$ of the reduced Berkovits theory can be derived from the improper field redefinition $F[\Psi_B]$ relating the reduced Berkovits theory to a free theory. Specifically, we prove the formula
\begin{equation}\tilde{B}_{n+1} = -\sum_{k=0}^n\sum_{j=0}^k (\omega^{j+1}\circ F_{k+1})(\mathbb{I}^{\otimes j}\otimes Q F_{n-k+1}\otimes\mathbb{I}^{k-j}).\label{eq:tildeB}\end{equation}
In the following derivation we will not assume that $\xi$ is in the small Hilbert space. Incidentally, the products $M_{n+1}$ of the $A_\infty$ theory can be computed with an analogous formula in terms of products $G_{n+1}$ of the cohomomorphism $\G$. However, since $\G$ is cyclic, this reduces to $\M=\G^{-1}\Q\G$. 

We start with the reduced Berkovits action
\begin{equation}S_B=-\frac{1}{g_o^2}\int_0^1 dt\,\left\langle \xi\PsiB,Q\Big((\eta e^{t\xi\PsiB}) e^{-t\xi\PsiB}\Big)\right\rangle_L .\end{equation}
We project the second entry of the BPZ inner product into the small Hilbert space by inserting $\eta\xi$, replace the large Hilbert space BPZ inner product with the small Hilbert space symplectic form, and substitute $F[\PsiB]$ to find
\begin{eqnarray}
S_B \lineup =  \frac{1}{g_o}\int_0^1 dt\,\omega_S(\PsiB,\eta\xi QF[t\PsiB]),\nonumber\\
\lineup = \frac{1}{g_o}\sum_{n=0}^{\infty} \frac{1}{n+2}\omega_S(\PsiB,\eta\xi QF_{n+1}(\underbrace{\PsiB,...,\PsiB}_{n+1\ \mathrm{times}})).
\end{eqnarray}
From this we can read off a sequence of degree odd non-cyclic products
\begin{eqnarray}
\Bnc_{n+2}\lineup = \frac{1}{g_o}[\eta,\xi QF_{n+2}],\nonumber\\
\lineup = QF_{n+2}+\sum_{k=0}^{n}\xi Q m_2(F_{k+1}\otimes F_{n-k+1}).
\end{eqnarray}
In the second step we used \eq{etaF} to compute the action of $\eta$ on $F_{n+2}$. The cyclic multi-string products $\tilde{B}_{n+2}$ are obtained by taking the cyclic projection:
\begin{equation}\tilde{B}_{n+2} = \frac{1}{n+3}\left(1+\sum_{j=0}^{n+1}\omega^{j+1}\right)\circ\Bnc_{n+2}.\end{equation}
Plugging in,
\begin{eqnarray}
\tilde{B}_{n+2}\lineup  = \frac{1}{n+3}\left[QF_{n+2}+\sum_{j=0}^{n+1}\omega^{j+1}\circ(QF_{n+2})\right.\nonumber\\
\lineup\ \ \ \ \ \ \ \ \ \ \left.+\sum_{k=0}^{n}\xi Q m_2(F_{k+1}\otimes F_{n-k+1})
+\sum_{k=0}^{n}\sum_{j=0}^{n+1}\omega^{j+1}\circ(\xi Q m_2(F_{k+1}\otimes F_{n-k+1}))\right].
\label{eq:tB1}
\end{eqnarray}
Computing the cyclic permutations we find 
\begin{equation}
\omega^{j+1} \circ(Q F_{n+2}) = -(\omega^{j+1}\circ F_{n+2})(\mathbb{I}^{\otimes j}\otimes Q\otimes\mathbb{I}^{\otimes n+1-j}) .
\end{equation}
for the second term, and either
\begin{eqnarray}
\omega^{j+1}\circ (\xi Q m_2(F_{k+1}\otimes F_{n-k+1}))\lineup = -(\omega^{j+1}\circ F_{n-k+1})(\mathbb{I}^{\otimes j}\otimes m_2(Q\xi\otimes F_{k+1})\otimes \mathbb{I}^{n-k-j}),\nonumber\\
\lineup \ \ \ \mathrm{for}\ 0\leq j\leq n-k,
\end{eqnarray}
or
\begin{eqnarray}
\omega^{(n-k+1)+j+1}\circ (\xi Q m_2(F_{k+1}\otimes F_{n-k+1}))\lineup = -(\omega^{j+1}\circ F_{k+1})(\mathbb{I}^{\otimes j}\otimes m_2(F_{n-k+1}\otimes Q\xi)\otimes \mathbb{I}^{k-j}),\nonumber\\
\lineup \ \ \ \mathrm{for}\ 0\leq j\leq k
\end{eqnarray}
for the fourth term. Then the fourth term in \eq{tB1} breaks into two pieces, depending on the number of cyclic permutations taken:
\begin{eqnarray}
\tilde{B}_{n+2}\lineup  = \frac{1}{n+3}\left[QF_{n+2}-\sum_{j=0}^{n+1}(\omega^{j+1}\circ F_{n+2})(\mathbb{I}^{\otimes j}\otimes Q\otimes \mathbb{I}^{n+1-j})+\sum_{k=0}^{n}\xi Q m_2(F_{k+1}\otimes F_{n-k+1})\right.\nonumber\\
\lineup\ \ \ \left.
+\sum_{k=0}^{n}\sum_{j=0}^{n-k}\omega^{j+1}\circ(\xi Q m_2(F_{k+1}\otimes F_{n-k+1}))
+\sum_{k=0}^{n}\sum_{j=0}^{k}\omega^{(n-k+1)+j+1}\circ(\xi Q m_2(F_{k+1}\otimes F_{n-k+1}))\right],\nonumber\\
\lineup = \frac{1}{n+3}\left[QF_{n+2}+\sum_{j=0}^{n+1}(\omega^{j+1}\circ F_{n+2})(\mathbb{I}^{\otimes j}\otimes Q\otimes \mathbb{I}^{n+1-j})+\sum_{k=0}^{n}\xi Q m_2(F_{k+1}\otimes F_{n-k+1})\right.\nonumber\\
\lineup\ \ \ \ \ \ \ \ \ \ \ 
-\sum_{k=0}^{n}\sum_{j=0}^{n-k}(\omega^{j+1}\circ F_{n-k+1})(\mathbb{I}^{\otimes j}\otimes m_2(Q\xi\otimes F_{k+1})\otimes \mathbb{I}^{n-k-j})
\nonumber\\
\lineup\ \ \ \ \ \ \ \ \ \ \ \left.-\sum_{k=0}^{n}\sum_{j=0}^{k}(\omega^{j+1}\circ F_{k+1})(\mathbb{I}^{\otimes j}\otimes m_2(F_{n-k+1}\otimes Q\xi)\otimes \mathbb{I}^{k-j})\right].
\end{eqnarray}
Relabeling indices allows us to combine the fourth and fifth terms:
\begin{eqnarray}
\tilde{B}_{n+2} \lineup = \frac{1}{n+3}\left[QF_{n+2}+\sum_{j=0}^{n+1}(\omega^{j+1}\circ F_{n+2})(\mathbb{I}^{\otimes j}\otimes Q\otimes \mathbb{I}^{n+1-j})+\sum_{k=0}^{n}\xi Q m_2(F_{k+1}\otimes F_{n-k+1})\right.
\nonumber\\
\lineup\ \ \ \ \ \ \ \ \ \ \ \left.-\sum_{k=0}^{n}\sum_{j=0}^{k}(\omega^{j+1}\circ F_{k+1})(\mathbb{I}^{\otimes j}\otimes m_2(F_{n-k+1}\otimes Q\xi+Q\xi\otimes F_{n-k+1})\otimes \mathbb{I}^{k-j})\right].
\end{eqnarray}
It will be convenient to use $[Q,m_2]=0$ to pull $Q$ away from the $\xi$ in the last two terms. Then we find:
\begin{eqnarray}
\tilde{B}_{n+2} \lineup = \frac{1}{n+3}\left[QF_{n+2}+\sum_{j=0}^{n+1}(\omega^{j+1}\circ F_{n+2})(\mathbb{I}^{\otimes j}\otimes Q\otimes \mathbb{I}^{n+1-j})\right.\nonumber\\
\lineup\ \ \ \ \ \ \ \ \ \ \ -\sum_{k=0}^{n}\xi m_2(QF_{k+1}\otimes F_{n-k+1})-\sum_{k=0}^{n}\xi m_2(F_{k+1}\otimes QF_{n-k+1})\nonumber\\
\lineup\ \ \ \ \ \ \ \ \ \ \
+\sum_{k=0}^{n}\sum_{j=0}^{k}(\omega^{j+1}\circ F_{k+1})(\mathbb{I}^{\otimes j}\otimes Qm_2(F_{n-k+1}\otimes\xi+\xi\otimes F_{n-k+1})\otimes \mathbb{I}^{k-j})
\nonumber\\
\lineup\ \ \ \ \ \ \ \ \ \ \ \left.+\sum_{k=0}^{n}\sum_{j=0}^{k}(\omega^{j+1}\circ F_{k+1})(\mathbb{I}^{\otimes j}\otimes m_2(QF_{n-k+1}\otimes \xi-\xi \otimes QF_{n-k+1})\otimes \mathbb{I}^{k-j})\right].
\end{eqnarray}
Using \eq{Frec} the fourth term simplifies, and reshuffling its position to the third term gives
\begin{eqnarray}
\tilde{B}_{n+2} \lineup = \frac{1}{n+3}\left[QF_{n+2}+\sum_{j=0}^{n+1}(\omega^{j+1}\circ F_{n+2})(\mathbb{I}^{\otimes j}\otimes Q\otimes \mathbb{I}^{n+1-j})\right.\nonumber\\
\lineup\ \ \ \ \ \ \ \ \ \ \ -\sum_{k=0}^{n}\sum_{j=0}^{k}(n-k+2)(\omega^{j+1}\circ F_{k+1})(\mathbb{I}^{\otimes j}\otimes QF_{n-k+2}\otimes \mathbb{I}^{k-j})\nonumber\\
\lineup\ \ \ \ \ \ \ \ \ \ \ -\sum_{k=0}^{n}\xi m_2(QF_{k+1}\otimes F_{n-k+1})-\sum_{k=0}^{n}\xi m_2(F_{k+1}\otimes QF_{n-k+1})
\nonumber\\
\lineup\ \ \ \ \ \ \ \ \ \ \ \left.+\sum_{k=0}^{n}\sum_{j=0}^{k}(\omega^{j+1}\circ F_{k+1})(\mathbb{I}^{\otimes j}\otimes m_2(QF_{n-k+1}\otimes \xi-\xi \otimes QF_{n-k+1})\otimes \mathbb{I}^{k-j})\right].\label{eq:tB22}
\end{eqnarray}
To simplify further we need to compute the cyclic permutations of equation \eq{Frec}:
\begin{equation}F_{m+2} = -\frac{1}{m+2}m_2(\xi\otimes F_{m+1}+F_{m+1}\otimes \xi).\end{equation}
We find
\begin{eqnarray}
\omega\circ F_{m+2} \lineup 
= \frac{1}{m+2}\Big((\omega\circ F_{m+1})(m_2(\mathbb{I}\otimes \xi)\otimes \mathbb{I}^{\otimes m}) +\xi m_2(\mathbb{I}\otimes F_{m+1})\Big),\nonumber\\
\omega^{j+1}\circ F_{m+2} \lineup 
= \frac{1}{m+2}\Big((\omega^{j+1}\circ F_{m+1})(\mathbb{I}^{\otimes j}\otimes m_2(\mathbb{I}\otimes \xi)\otimes \mathbb{I}^{\otimes m-j}) 
\nonumber\\
\lineup\ \ \ \ \ \ \ \ \ \ \ \ \ +(\omega^j\circ F_{m+1})(\mathbb{I}^{\otimes j-1}\otimes m_2(\xi\otimes\mathbb{I})\otimes\mathbb{I}^{m-j})\Big)
,\ \ \ \ \ (1\leq j\leq m),\nonumber\\
\omega^{m+2}\circ F_{m+2} \lineup 
= \frac{1}{m+2}\Big(\xi m_2(F_{m+1}\otimes\mathbb{I})+(\omega^{m+1}\circ F_{m+1})(\mathbb{I}^{\otimes m}\otimes m_2(\xi\otimes\mathbb{I}))\Big)
.\label{eq:cycrec}
\end{eqnarray}
Let us focus on the last three terms in \eq{tB22}:
\begin{eqnarray}
\mathrm{last\ }3\ \mathrm{terms\ in\ \eq{tB22}}\lineup = -\sum_{k=0}^{n}\xi m_2(QF_{k+1}\otimes F_{n-k+1})-\sum_{k=0}^{n}\xi m_2(F_{k+1}\otimes QF_{n-k+1})
\nonumber\\
\lineup\ \ \ +\sum_{k=0}^{n}\sum_{j=0}^{k}(\omega^{j+1}\circ F_{k+1})(\mathbb{I}^{\otimes j}\otimes m_2(QF_{n-k+1}\otimes \xi-\xi \otimes QF_{n-k+1})\otimes \mathbb{I}^{k-j}),\nonumber\\
\lineup \!\!\!\!\!\!\!\!\!\!\!\!\!\!\!\!\!\!\!\!\!\!\!\!\!\!\!\!\!\!
= -\sum_{k=0}^{n}\xi m_2(\mathbb{I}\otimes F_{k+1})(QF_{n-k+1}\otimes\mathbb{I}^{\otimes k+1})
-\sum_{k=0}^{n}\xi m_2(F_{k+1}\otimes \mathbb{I})(\mathbb{I}^{\otimes k+1}\otimes QF_{n-k+1})
\nonumber\\
\lineup \!\!\!\!\!\!\!\!\!\!\!\!\!\!\!\!\!\!\!\!\!\!\!\!\!\!\!\!\!\!\ \ \ 
-\sum_{k=0}^{n}\sum_{j=0}^{k}(\omega^{j+1}\circ F_{k+1})(\mathbb{I}^{\otimes j}\otimes m_2(\mathbb{I}\otimes \xi)\otimes\mathbb{I}^{\otimes k-j})(\mathbb{I}^{\otimes j}\otimes QF_{n-k+1}\otimes \mathbb{I}^{\otimes k-j+1})
\nonumber\\
\lineup\!\!\!\!\!\!\!\!\!\!\!\!\!\!\!\!\!\!\!\!\!\!\!\!\!\!\!\!\!\!\ \ \ 
-\sum_{k=0}^{n}\sum_{j=1}^{k+1}(\omega^{j}\circ F_{k+1})(\mathbb{I}^{\otimes j-1}\otimes m_2(\xi\otimes \mathbb{I})\otimes\mathbb{I}^{\otimes k+1-j})(\mathbb{I}^{\otimes j}\otimes QF_{n-k+1}\otimes \mathbb{I}^{\otimes k-j+1}),
\end{eqnarray}
where we relabeled indices $k\to n-k$ in the first term and $j\to j-1$ in the last term, and factored out $QF_{n-k+1}$. Now the last two terms above can be combined except for a mismatch at the extremes of summation over $j$:
\begin{eqnarray}
\mathrm{last\ }3\ \mathrm{terms\ in\ \eq{tB22}} \lineup =\nonumber\\
\lineup\!\!\!\!\!\!\!\!\!\!\!\!\!\!\!\!\!\!\!\!\!\!\!\!\!\!\!\!\!\!
-\sum_{k=0}^{n}\xi m_2(\mathbb{I}\otimes F_{k+1})(QF_{n-k+1}\otimes\mathbb{I}^{\otimes k+1})
-\sum_{k=0}^{n}\xi m_2(F_{k+1}\otimes \mathbb{I})(\mathbb{I}^{\otimes k+1}\otimes QF_{n-k+1})\nonumber\\
\lineup \!\!\!\!\!\!\!\!\!\!\!\!\!\!\!\!\!\!\!\!\!\!\!\!\!\!\!\!\!\!
-\sum_{k=0}^{n}(\omega\circ F_{k+1})(m_2(\mathbb{I}\otimes \xi)\otimes\mathbb{I}^{\otimes k})(QF_{n-k+1}\otimes \mathbb{I}^{\otimes k+1})\nonumber\\
\lineup\!\!\!\!\!\!\!\!\!\!\!\!\!\!\!\!\!\!\!\!\!\!\!\!\!\!\!\!\!\!
-\sum_{k=0}^{n}(\omega^{k+1}\circ F_{k+1})(\mathbb{I}^{\otimes k}\otimes m_2(\xi\otimes \mathbb{I}))(\mathbb{I}^{\otimes k+1}\otimes QF_{n-k+1})
\nonumber\\
\lineup \!\!\!\!\!\!\!\!\!\!\!\!\!\!\!\!\!\!\!\!\!\!\!\!\!\!\!\!\!\!
-\sum_{k=0}^{n}\sum_{j=1}^{k}\Bigg[(\omega^{j+1}\circ F_{k+1})(\mathbb{I}^{\otimes j}\otimes m_2(\mathbb{I}\otimes \xi)\otimes\mathbb{I}^{\otimes k-j})
\nonumber\\
\lineup \!\!\!\!\!\!\!\!\!\!\!\!\!\!\!\!
+(\omega^{j}\circ F_{k+1})(\mathbb{I}^{\otimes j-1}\otimes m_2(\xi\otimes \mathbb{I})\otimes\mathbb{I}^{\otimes k+1-j})\Bigg](\mathbb{I}^{\otimes j}\otimes QF_{n-k+1}\otimes \mathbb{I}^{\otimes k-j+1}).
\end{eqnarray}
Now look at the terms where $Q F_{n-k+1}$ appears at the same position on the $n+2$ string input. Comparing to \eq{cycrec} we see that the $\xi$s and $m_2$s can be absorbed into cyclic permutations of $F_{k+2}$, so we arrive at a simplification
\begin{eqnarray}
\mathrm{last\ }3\ \mathrm{terms\ in\ \eq{tB22}} \lineup =\nonumber\\
\lineup\!\!\!\!\!\!\!\!\!\!\!\!\!\!\!\!\!\!\!\!\!\!\!\!\!\!\!\!\!\!
-\sum_{k=0}^{n}(k+2)(\omega\circ F_{k+2})(QF_{n-k+1}\otimes\mathbb{I}^{\otimes k+1})
-\sum_{k=0}^{n}(k+2)(\omega^{k+2}\circ F_{k+2})(\mathbb{I}^{\otimes k+1}\otimes QF_{n-k+1})
\nonumber\\
\lineup \!\!\!\!\!\!\!\!\!\!\!\!\!\!\!\!\!\!\!\!\!\!\!\!\!\!\!\!\!\!
-\sum_{k=0}^{n}\sum_{j=1}^{k}(k+2)(\omega^{j+1}\circ F_{k+2})(\mathbb{I}^{\otimes j}\otimes QF_{n-k+1}\otimes \mathbb{I}^{\otimes k-j+1})
,\nonumber\\
\lineup \!\!\!\!\!\!\!\!\!\!\!\!\!\!\!\!\!\!\!\!\!\!\!\!\!\!\!\!\!\!\!\!\!\!\!\!\!\!
=-\sum_{k=0}^{n}\sum_{j=0}^{k+1}(k+2)(\omega^{j+1}\circ F_{k+2})(\mathbb{I}^{\otimes j}\otimes QF_{n-k+1}\otimes \mathbb{I}^{\otimes k-j+1}).
\end{eqnarray}
Therefore \eq{tB22} simplifies to
\begin{eqnarray}
\tilde{B}_{n+2} \lineup = \frac{1}{n+3}\left[QF_{n+2}+\sum_{j=0}^{n+1}(\omega^{j+1}\circ F_{n+2})(\mathbb{I}^{\otimes j}\otimes Q\otimes \mathbb{I}^{n+1-j})\right.\nonumber\\
\lineup\ \ \ \ \ \ \ \ \ \ \ -\sum_{k=0}^{n}\sum_{j=0}^{k}(n-k+2)(\omega^{j+1}\circ F_{k+1})(\mathbb{I}^{\otimes j}\otimes QF_{n-k+2}\otimes \mathbb{I}^{k-j})\nonumber\\
\lineup\ \ \ \ \ \ \ \ \ \ \ \left.
-\sum_{k=0}^{n}\sum_{j=0}^{k+1}(k+2)(\omega^{j+1}\circ F_{k+2})(\mathbb{I}^{\otimes j}\otimes QF_{n-k+1}\otimes \mathbb{I}^{\otimes k-j+1})\right].
\end{eqnarray}
In the last term relabel $k\to k-1$
\begin{eqnarray}
\tilde{B}_{n+2} \lineup = \frac{1}{n+3}\left[QF_{n+2}+\sum_{j=0}^{n+1}(\omega^{j+1}\circ F_{n+2})(\mathbb{I}^{\otimes j}\otimes Q\otimes \mathbb{I}^{n+1-j})\right.\nonumber\\
\lineup\ \ \ \ \ \ \ \ \ \ \ -\sum_{k=0}^{n}\sum_{j=0}^{k}(n-k+2)(\omega^{j+1}\circ F_{k+1})(\mathbb{I}^{\otimes j}\otimes QF_{n-k+2}\otimes \mathbb{I}^{k-j})\nonumber\\
\lineup\ \ \ \ \ \ \ \ \ \ \ \left.
-\sum_{k=1}^{n+1}\sum_{j=0}^{k}(k+1)(\omega^{j+1}\circ F_{k+1})(\mathbb{I}^{\otimes j}\otimes QF_{n-k+2}\otimes \mathbb{I}^{\otimes k-j})\right].
\end{eqnarray}
The third and fourth terms can now be combined except for a mismatch at the extremes of summation over $k$:
\begin{eqnarray}
\tilde{B}_{n+2} \lineup = \frac{1}{n+3}\left[QF_{n+2}+\sum_{j=0}^{n+1}(\omega^{j+1}\circ F_{n+2})(\mathbb{I}^{\otimes j}\otimes Q\otimes \mathbb{I}^{n+1-j})\right.\nonumber\\
\lineup\ \ \ \ \ \ \ \ \ \ \ -\sum_{k=1}^{n}\sum_{j=0}^{k}(n+3)(\omega^{j+1}\circ F_{k+1})(\mathbb{I}^{\otimes j}\otimes QF_{n-k+2}\otimes \mathbb{I}^{k-j})-(n+2)(\omega\circ F_{1})(QF_{n+2}),\nonumber\\
\lineup\ \ \ \ \ \ \ \ \ \ \ \left. -\sum_{j=0}^{n+1}(n+2)(\omega^{j+1}\circ F_{n+2})(\mathbb{I}^{\otimes j}\otimes QF_{1}\otimes \mathbb{I}^{\otimes n+1-j})\right].
\end{eqnarray}
Noting $F_1=\mathbb{I}$ and $\omega\circ \mathbb{I} = -\mathbb{I}$, the first and second terms can be combined with the fourth and fifth terms, which allows us to cancel the overall factor of $n+3$. We can then absorb the terms into a single summation, finding 
\begin{equation}
\tilde{B}_{n+2} = -\sum_{k=0}^{n+1}\sum_{j=0}^k (\omega^{j+1}\circ F_{k+1})(\mathbb{I}^{\otimes j}\otimes Q F_{n-k+2}\otimes\mathbb{I}^{k-j}),\end{equation}
which reproduces \eq{tildeB}.

\end{appendix}

\end{document}